\RequirePackage[l2tabu, orthodox]{nag}
\RequirePackage{snapshot}

\documentclass[journal,draftcls,onecolumn]{IEEEtran}

\usepackage{amsmath,amssymb}
\usepackage{url,booktabs,bm}
\usepackage{mathtools}
\usepackage{enumerate}

\usepackage{multirow}
\usepackage{siunitx}
\usepackage{textcomp}
\usepackage{cite}

\usepackage{graphicx}
\usepackage{subfigure}
\usepackage{balance,xcolor}
\usepackage{algorithm,algpseudocode}

\usepackage{xr} 
\usepackage{hyperref,pdfcomment}

\usepackage[utf8]{inputenc}

\graphicspath{{../}{../ImgFig/}}

\newtheorem{LEO}{LEO}
\newtheorem{PEO}{PEO}
\newtheorem{Lemma}[LEO]{Lemma}
\newtheorem{Proposition}[PEO]{Proposition}

\begin{document}

\title{%
Detecting Changes in Fully Polarimetric SAR Imagery with Statistical Information Theory}

\author{Abra\~ao~D.~C.~Nascimento,
        Alejandro~C.~Frery,~\IEEEmembership{Senior Member}, 
        and
        Renato~J.~Cintra,~\IEEEmembership{Senior Member}
				}

\markboth{
Published in the IEEE TGRS}
{Nascimento \MakeLowercase{\emph{et~al.}}: Nascimento}

\maketitle
\begin{abstract}
Images obtained from coherent illumination processes are contaminated with speckle. 
A prominent example of such imagery systems is the polarimetric synthetic aperture radar (PolSAR).
For such remote sensing tool the speckle interference pattern appears in the form of a positive definite Hermitian matrix, 
which requires specialized models 
and makes change detection a hard task.
The scaled complex Wishart distribution is a widely used model for PolSAR images.
Such distribution is defined by two parameters: the number of looks and  the complex covariance matrix.
The last parameter contains all the necessary information to characterize the backscattered data and, thus, identifying changes in a sequence of images can be formulated as a problem of verifying whether the complex covariance matrices differ at two or more takes.
This paper proposes a comparison between a classical change detection method based on the likelihood ratio
and three statistical methods
that depend on information-theoretic measures: the Kullback-Leibler distance and two entropies.
The performance of these four tests was quantified in terms of their sample test powers and sizes using simulated data.
The tests are then applied to actual PolSAR data.
The results provide evidence that tests based on entropies may outperform those based on the Kullback-Leibler distance and likelihood ratio statistics.
\end{abstract}
\begin{IEEEkeywords}
contrast, information theory, Wishart, hypothesis test, change detection.
\end{IEEEkeywords}

\section{Introduction}\label{conparison:intro}

Synthetic aperture radar (SAR) has been widely used as an important system for information extraction in remote sensing applications.
Such microwave active sensors have as main advantages the following features:
(i)~their operation is not determined by day time, neither weather conditions
and
(ii)~they are capable of providing high spatial image resolution.

In recent years, the interest in understanding such type of imagery in a multidimensional and multilook perspective has increased. 
Such systems are called ``polarimetric SAR'' (PolSAR).
In this case, obtaining of PolSAR data obeys the following dynamic: a scene is mapped with polarized pulses which are backscattered by the scene and captured by a sensor to form an image. 
As a result, PolSAR measurements record the amplitude and phase of backscattered signals for possible combinations of linear reception and transmission polarizations: HH, HV, VH, and VV (H for horizontal and V for vertical polarization).

However, since the acquired images stem from a coherent illumination process, they are affected by a signal-dependent granular noise called ``speckle''~\cite{LeePottier2009PolarimetricRadarImaging}. 
Such noise has a multiplicative nature and its intensity does not follow the Gaussian law. 
Thus, analyzing PolSAR images requires tailored image processing based on the statistical properties of speckled data.

PolSAR theory prescribes that the returned (backscattered) signal of distributed targets is adequately represented by its complex covariance matrix.
Under the assumption that the complex scattering coefficients are jointly circular Gaussian, the Wishart distribution is the statistical model for multilook PolSAR data. 
This paper adopts the assumption that a PolSAR image is well described by such distribution.

Change detection methods aim at identifying differences in the scene configuration at distinct observation instants.
Such procedures have achieved a prominent position in recent decades~\cite{LuMauseBrondMoran2004}.
Indeed, literature reports several approaches for change detection problems, among them:
\begin{enumerate}[(i)]
\item image ratioing~\cite{Rignot1993,OliverandQuegan1998,Liuetal2015,MoserSerpicoVernazza2007},
\item multitemporal coherence analysis~\cite{LL2015},
\item spatiotemporal contextual classification~\cite{Melgani20021053,Atto2013},
\item\label{list:LR} Hotelling-Lawley and likelihood ratio tests~\cite{SilvaCribariFrery:ImprovedLikelihood:Environmetrics,Conradsen2003,Akbari2013,Liu2014,C2015,Carotenuto2015,Carotenuto2015L,Meng2015,Nielsen2015,Conradsen2016} and robust tests~\cite{LinPerissin2017},
\item combination of image ratioing and the generalized minimum-error method~\cite{MoserandSerpico2006}, 
\item detection algorithms based on Lagrange optimization~\cite{Marino2014},
\item\label{list:IT} information-theoretic measures for change detection~\cite{IngladaMercier2007,
ConditionalCopulasChangeDetection,HypothesisTestingSpeckledDataStochasticDistances,FreryCintraNascimento2012,FreryCintraNascimento2014,Marino2013,Atto2013, Zheng2014,Ratha2017}
and
\item change detection with post-classification~\cite{Akbarietal2004}. 
\end{enumerate}
This paper advances points~(\ref{list:LR}) and~(\ref{list:IT}) above.

The change detection process is theoretically rooted in the hypothesis test theory and the proposal of statistical similarity measures~\cite{Radkeetal2005}. 
In particular, hypothesis tests based on the complex covariance matrix have been sought for PolSAR data analysis. 
Many statistical approaches have been developed in order to reach this goal.

Conradsen~\emph{et~al.}~\cite{Conradsen2003} proposed a methodology based on the likelihood ratio test defined by two random samples from the complex Wishart distribution. 
Subsequently, this technique was applied to edge detection in PolSAR images by Schou~\emph{et~al.}~\cite{Schou2003}.
Recently, 
Conradsen~\emph{et~al.}~\cite{Conradsen2016} 
extended likelihood-based detection for 
PolSAR time series.
Kersten and Ainsworth~\cite{KerstenLeeAinsworth} compared three test statistics (the contrast ratio, ellipticity, and Bartlett tests).
It was found that the method based on the contrast ratio is more robust to variations in the covariance estimates on actual data.
In a complementary study, Molinier and Rauste~\cite{MolinierRauste2007} compared six polarimetric change detection methods.
As a conclusion, the methods directly derived from the Wishart distribution outperformed other approaches
as they provide explicit thresholds.
Recently, Akbari~\emph{et~al.}~\cite{Akbarietal2016} proposed a change detector involving the Hotelling-Lawley trace (HLT) which, asymptotically, follows the Fisher-Snedecor distribution.
The authors provided evidence that 
the HLT test may outperform the Bartlett test in some scenarios.

Several works have employed information-theoretic tools as a pre-processing step for change detection in PolSAR images.
They can be categorized into two approaches: 
one is based only on discrimination measures,
whereas the other considers the asymptotic distribution of such tools.

In the first category, Inglada and Mercier~\cite{IngladaMercier2007} proposed a new similarity measure for automatic change detection in multitemporal SAR images.
Such measure was derived considering the symmetrized Kullback-Leibler (KL) divergence (or distance) between the Edgeworth series expansions for two distinct elements of the $\mathcal{K}$ distribution from the Pearson System~\cite{Ward1981} for intensity SAR data.
In~\cite{ConditionalCopulasChangeDetection}, the KL measure is improved by means of copula-based quantile regression to generate local change measures.
Further, Erten~\emph{et~al.}~\cite{Ertenetal2012} proposed a new method based on mutual information for quantifying the coherent similarity between temporal multichannel PolSAR images.
Atto~\emph{et~al.}~\cite{Atto2013} used the KL divergence for spatio-temporal change detection in image time series.

In the second category, Nascimento~\emph{et~al.}~\cite{HypothesisTestingSpeckledDataStochasticDistances} derived hypothesis tests based on several distance measures between ${\mathcal G}^0$ distributions~\cite{freryetal1997a}. 
In terms of the nature of the image data, these results were extended in~\cite{FreryCintraNascimento2012,FreryCintraNascimento2014} and applied to boundary detection~\cite{Nascimentoetaieee12013} and filtering~\cite{TorresPolarimetricFilterPatternRecognition} in PolSAR images.  
All these references derived new proposals using contrast measures
designed from the scaled complex Wishart law.
Recently,
Akbari~\emph{et~al.}~\cite{Akbarietal20152}
introduced
a change detector with
the HLT statistics as
the contrast measure based on the relaxed scaled Wishart likelihood.

This paper proposes three new change detection methodologies for fully polarimetric data.
Additionally, a new expression for the likelihood ratio statistics obtained from the scaled Wishart distribution is achieved, 
and its relationship with the individual distributions of the intensity channels is discussed.    
Using Monte Carlo simulation, we quantify the performance of four parametric methodologies for detecting the change: two considering Shannon and R\'enyi entropies, one stemming from the Kullback-Leibler distance, and one based on the classic likelihood ratio statistics.
The methods are compared by their empirical test size and power.
Finally, two experiments with actual PolSAR data are performed. 
Results provide evidence that the methods based on entropies
are superior.

This paper is organized as follows.
Section~\ref{comparison:model} provides the background of the statistical modeling.
A brief survey on parametric methodologies for hypothesis testing on complex covariance matrices is provided in Section~\ref{comparison:methodology}.
In Section~\ref{comparison:results}, we present a comparative study of change detection methods by means of Monte Carlo simulation.
Additionally, we perform two experiments with actual PolSAR data. 
Section~\ref{comparison:conclusion} summarizes the main results.

\section{Statistical Modeling for PolSAR Data}\label{comparison:model}

PolSAR systems represent each resolution cell by $p$ polarization elements comprising a complex random vector:
\begin{equation}
\label{backscattervectorr}
\boldsymbol{y}=[S_1\; S_2\;\cdots\; S_p]^\top
,
\end{equation}
where the superscript ${}^\top$ is the vector transposition.
In single-look PolSAR image processing,
$\boldsymbol{y}$ is admitted to obey the multivariate complex circular Gaussian distribution with zero mean~\cite{Goodmanb} whose probability density function (pdf) is:
$$
f_{\boldsymbol{y}}(\dot{\boldsymbol{y}};\boldsymbol{\Sigma})=\frac{1}{\pi^p|\boldsymbol{\Sigma}|}\exp\bigl(-\dot{\boldsymbol{y}}^{*}\boldsymbol{\Sigma}^{-1}\dot{\boldsymbol{y}}\bigr),
$$
where $\dot{\boldsymbol{y}}$ is an outcome of $\boldsymbol{y}$,  $|\cdot|$ is the matrix determinant, the superscript ${}^*$ denotes the complex conjugate transpose of a vector, $\boldsymbol{\Sigma}$ is the covariance matrix of  $\boldsymbol{y}$ such that $\boldsymbol{\Sigma}=\operatorname{E}\{\boldsymbol{y}\boldsymbol{y}^{*}\} $, and $\operatorname{E}\{\cdot\}$ is the statistical expectation operator.
This distribution is denoted by $\boldsymbol{y}\sim\mathcal{N}^{C}(\bm 0,\boldsymbol{\Sigma})$.
Besides being Hermitian and positive definite, $\boldsymbol{\Sigma}$ contains all the necessary information to characterize the backscattered data~\cite{Lopez-MartinezFabregasPottier2005}.

In order to improve the signal-to-noise ratio, $L$ independent and identically distributed samples are usually averaged in order to form the $L$-looks covariance matrix~\cite{EstimationEquivalentNumberLooksSAR}:	
$$
\boldsymbol{Z}=\frac{1}{L}\sum_{i=1}^L \boldsymbol{y}_i\boldsymbol{y}_i^{*},
$$
where $\boldsymbol{y}_i$, $i=1,2,\ldots,L$ are realizations of~\eqref{backscattervectorr}.
Under the aforementioned hypotheses, $\boldsymbol{Z}$ follows a scaled complex Wishart distribution. 
Having $\boldsymbol{\Sigma}$ and $L$ as parameters, such law is characterized by the following pdf:
\begin{equation}
 f_{\boldsymbol{Z}}(\dot{\boldsymbol{Z}};\boldsymbol{\Sigma},L) = \frac{L^{pL}|\dot{\boldsymbol{Z}}|^{L-p}}{|\boldsymbol{\Sigma}|^L \Gamma_p(L)} \exp\bigl[
-L\operatorname{tr}\bigl(\boldsymbol{\Sigma}^{-1}\dot{\boldsymbol{Z}}\bigr)\bigr],
\label{eq:denswishart}
\end{equation}
where $\Gamma_p(L)=\pi^{p(p-1)/2}\prod_{i=0}^{p-1}\Gamma(L-i)$, $L\geq p$, $\Gamma(\cdot)$ is the gamma function, and $\operatorname{tr}(\cdot)$ is the trace operator.
We denote it by $\boldsymbol{Z}\sim \mathcal{W}(\boldsymbol{\Sigma},L)$.
This distribution satisfies $\operatorname{E}\{\boldsymbol{Z}\}=\boldsymbol{\Sigma}$, which is a Hermitian positive definite matrix~\cite{EstimationEquivalentNumberLooksSAR}.
In practice, $L$ is treated as a parameter and must be estimated.
The resulting distribution is the relaxed Wishart distribution, and it is denoted by $\mathcal{W_R}(\boldsymbol{\Sigma},L)$~\cite{AnfinsenJenssenEltoft2009}. 

Due to its optimal asymptotic properties, 
we employ the maximum likelihood (ML) approach to estimate the parameters $\boldsymbol{\Sigma}$ and the equivalent number of looks~$L$.
Let $\underline{\boldsymbol{Z}}=\{\boldsymbol{Z}_1,\boldsymbol{Z}_2,\ldots,\boldsymbol{Z}_N\}$ be a random sample of size $N$ obtained from $\boldsymbol{Z}\sim \mathcal{W_R}(\boldsymbol{\Sigma},L)$. 
Setting $\ell_k(\boldsymbol{\theta})=\log f_{\boldsymbol{Z}}(\boldsymbol{Z}_k;\boldsymbol{\Sigma},L)$ for $\boldsymbol{\theta}=[\mathrm{vec}(\boldsymbol{\Sigma})^\top,\,L]^\top$ as the log-likelihood of the $k$th random matrix, $\boldsymbol{Z}_k$, from $\underline{\boldsymbol{Z}}$, solving $N^{-1}\sum_{k=1}^N \nabla \ell_k(\widehat{\boldsymbol{\theta}})=\boldsymbol{0}$, we have that
$
\widehat{\boldsymbol{\Sigma}}={N}^{-1}\sum_{k=1}^N {\boldsymbol{Z}_k},
$
and
\begin{align}
p\log\widehat{L}+\frac{1}{N}\sum_{k=1}^N \log|\boldsymbol{Z}_k|-\log|\widehat{\boldsymbol{\Sigma}}|-\psi_p^{(0)}(\widehat{L})=0,
\label{eqscore1}
\end{align}
where $\mathrm{vec}(\cdot)$ is the vectorization operator, 
$\psi_p^{(0)}(\cdot)$ is the zero order term of the $v$th-order multivariate polygamma function:
$$
\psi_p^{(v)}(L)=\sum_{i=0}^{p-1} \psi^{(v)}(L-i),
$$
and $\psi^{(v)}(\cdot)$ is the ordinary polygamma function expressed by
$$
\psi^{(v)}(L)=\frac{\partial^{v+1} \log\Gamma(L)}{\partial L^{v+1}},
$$
for $v\geq 0$; note that $\psi^{(0)}$ is the digamma function~\cite{AbramowitzStegun1994}.

Thus, the ML estimator of $\boldsymbol{\Sigma}$ is the sample mean, while $\widehat L$ is obtained
by solving the system shown in~\eqref{eqscore1}. 
We used the Newton-Raphson iterative method~\cite{gentle2002elements} to solve it.
The work by Anfinsen~\emph{et~al.}~\cite{EstimationEquivalentNumberLooksSAR} is an important reference on how to efficiently estimate $L$.

\begin{figure}[hbt]
\centering
\includegraphics[width=1\linewidth]{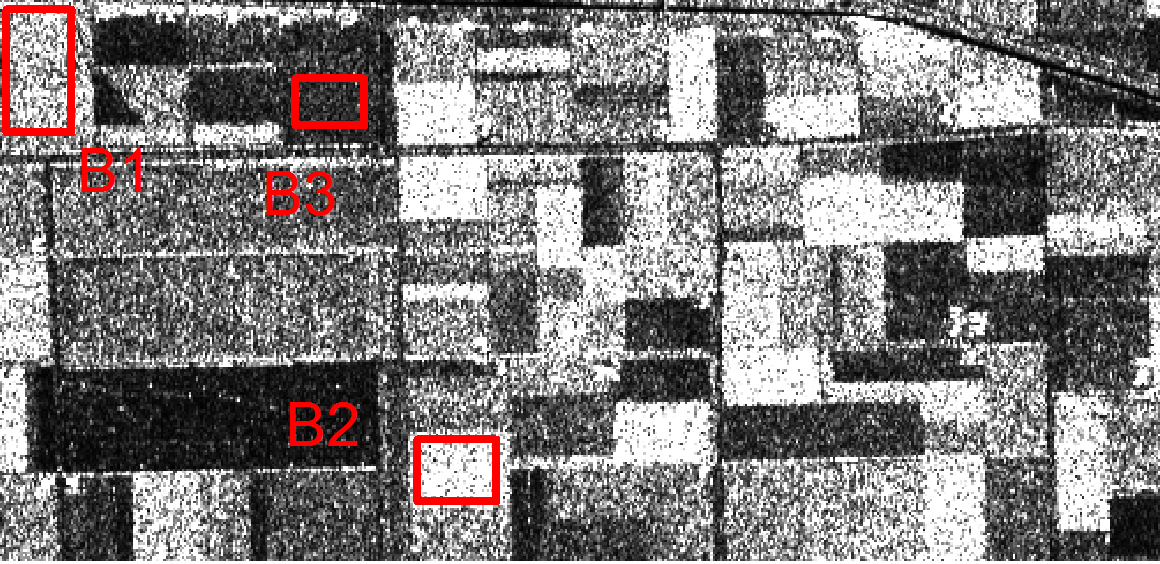}
\caption{AIRSAR image of Flevoland (channel HH).}
\label{Apli10}
\end{figure}

Fig.~\ref{Apli10} shows an area from the AIRSAR image of Flevoland, the Netherlands, obtained on August 1989~\cite{Doulgerisetal2011} with four nominal looks.
We delimited three regions of interest.

Table~\ref{tabelapplica} lists the ML parameter estimates as well as the sample sizes.
Each sample is taken from a single class without evidence of texture.
Notice that the estimates for the equivalent number of looks are very close, although lower than the nominal value.
We also show the determinant of the estimated covariance matrix.
This quantity, called \emph{geometric intensity} in~\cite{Skrunesetal2014}, is the \emph{generalized variance} in multivariate analysis; 
it can be used as a measure of mean backscatter~\cite{Goodmana}.
According to it, region B$_2$ presents the highest return, followed by B$_1$ and by B$_3$; this is in agreement with what is observed in channel HH, cf.\ Fig.~\ref{Apli10}.

\begin{table}[hbt]
\centering   
\caption{Estimated parameters on PolSAR data from Flevoland}\label{tabelapplica}
\begin{tabular}{c r@{.}l r@{}l  r} \toprule
Regions & \multicolumn{2}{c}{$\widehat{L}$} & \multicolumn{2}{c}{$|\widehat{\boldsymbol{\Sigma}}|$}  & \# pixels  \\ \midrule
$\text{B}_1$ &      $3$&$470$ &  $7.78$&$\times 10^{-8}$   &   $1566$   \\  
$\text{B}_2$ &      $3$&$514$ &  $9.45$&$\times 10^{-7}$    &   $980$   \\
$\text{B}_3$ &      $3$&$530$  &  $7.22$&$\times 10^{-10}$   &   $651$    \\  
\bottomrule
\end{tabular}
\end{table}

\begin{figure}[hbt]
\centering
\subfigure[Region B$_1$ \label{Apli11}]{\includegraphics[width=.48\linewidth]{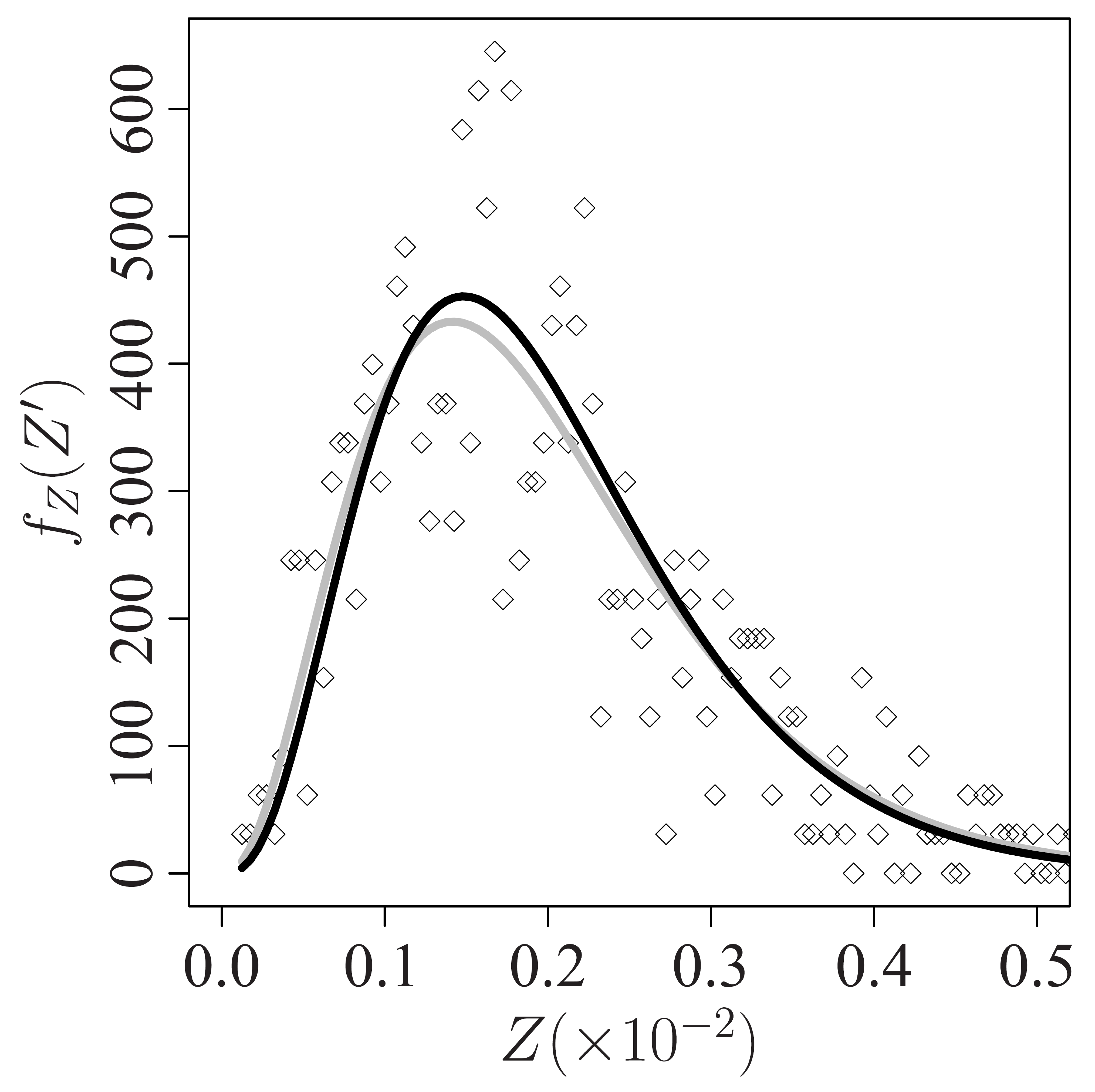}}
\subfigure[Region B$_2$\label{Apli12}]{\includegraphics[width=.48\linewidth]{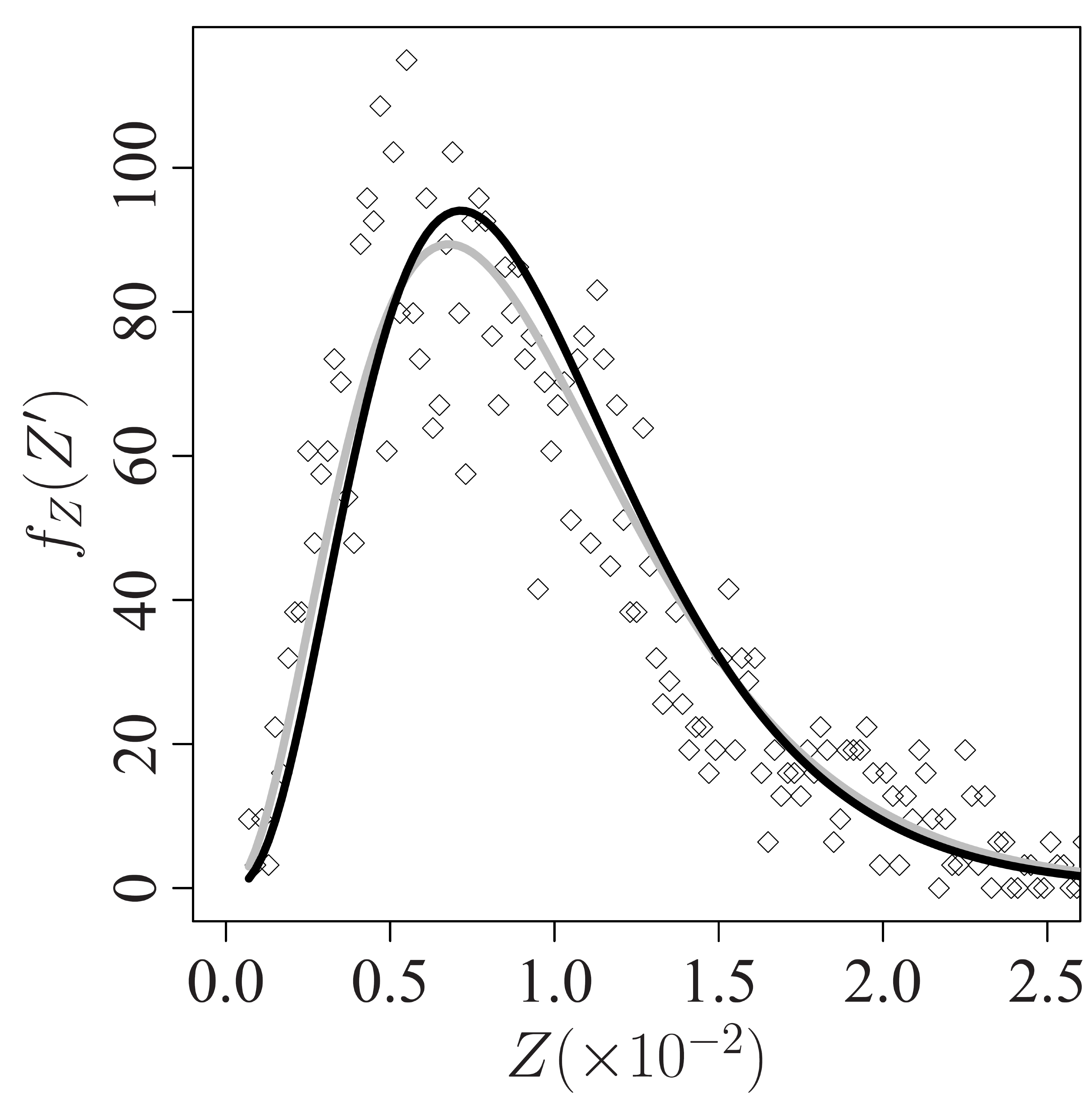}}
\subfigure[Region B$_3$\label{Apli13}]{\includegraphics[width=.48\linewidth]{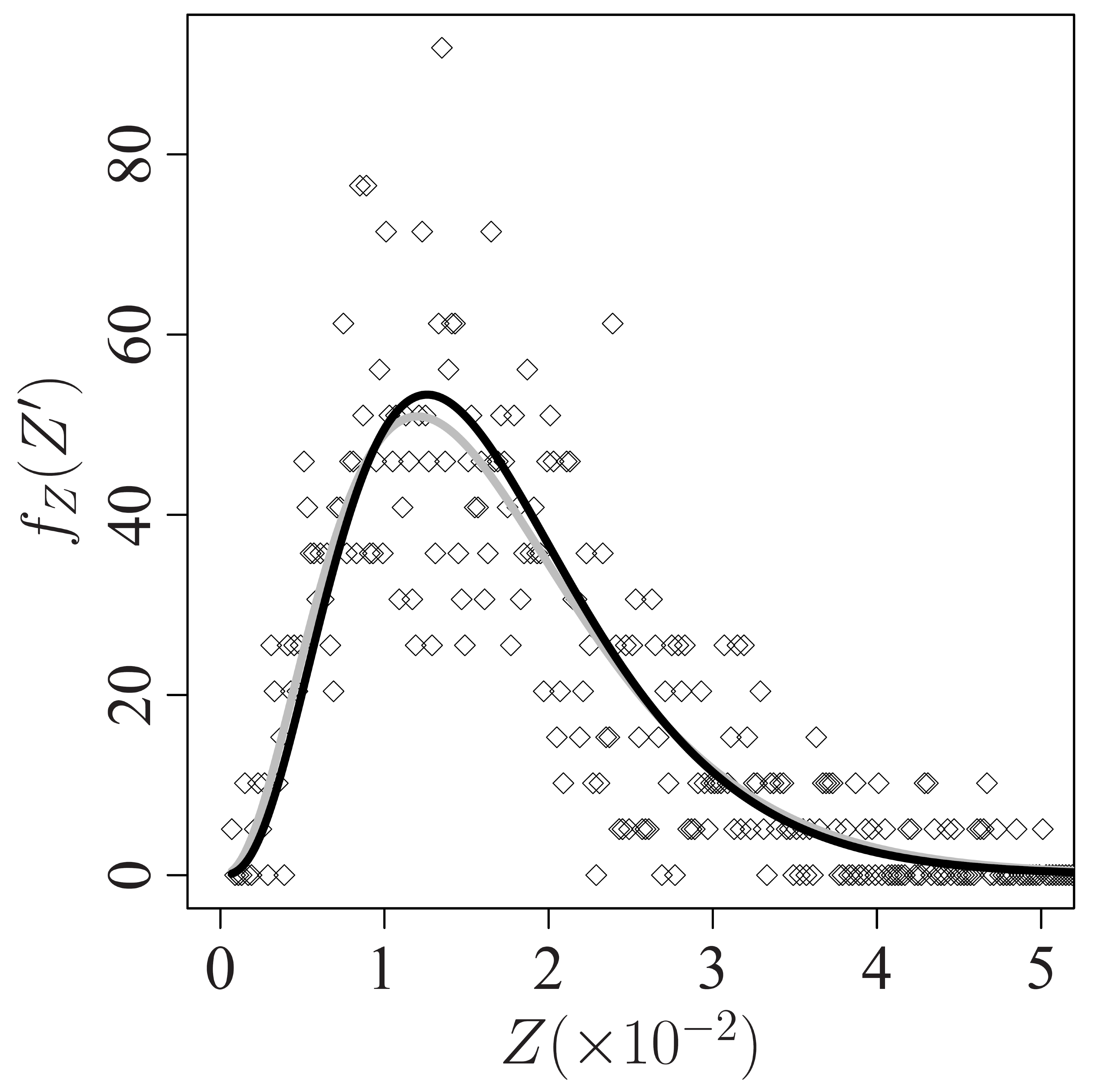}}
\caption{Histograms of HH channel data and densities with estimated number of looks (black) and fixed \textit{a priori} (gray), respectively.} 
\label{Apli1}  
\end{figure}

Fig.~\ref{Apli1} depicts the empirical densities of data from the agricultural regions along with the fitted marginal densities.
The scaled Wishart density collapses to the Gamma density:
\begin{equation}
f_{Z_i}(z_i;\theta_i,L)=\frac{L^L {z_i}^{L-1}}{\Gamma(L)\,\theta_i^L} \exp\bigl[-L\,\theta_i^{-1}z_i\bigr],
\label{singlepol}
\end{equation}
where $i\in\{\text{HH,HV,VV}\}$, $\theta_k$ is the element $(k,k)$ of $\boldsymbol{\Sigma}$, and $Z_k$ is the $(k,k)$-th entry of $\boldsymbol{Z}$.
In practice, $\theta_i$ represents the mean polarization channel $i\in\{1 (\text{HH}), 2 (\text{HV}), 3 (\text{VV})\}$.
Figs.~\ref{Apli11}-\ref{Apli13} show the data and the densities for the estimated number of looks
$\mathcal{W_R}(\widehat{\boldsymbol{\Sigma}},\widehat L)$ (black curve) and the fixed value
$\mathcal W(\widehat{\boldsymbol{\Sigma}},4)$ (gray curve).
These densities are remarkably close, and also to the histograms, so the Gamma assumption is reasonable.

\begin{figure}[hbt]
\centering
\subfigure[Region B1\label{la1}]{
\includegraphics[width=.48\linewidth]{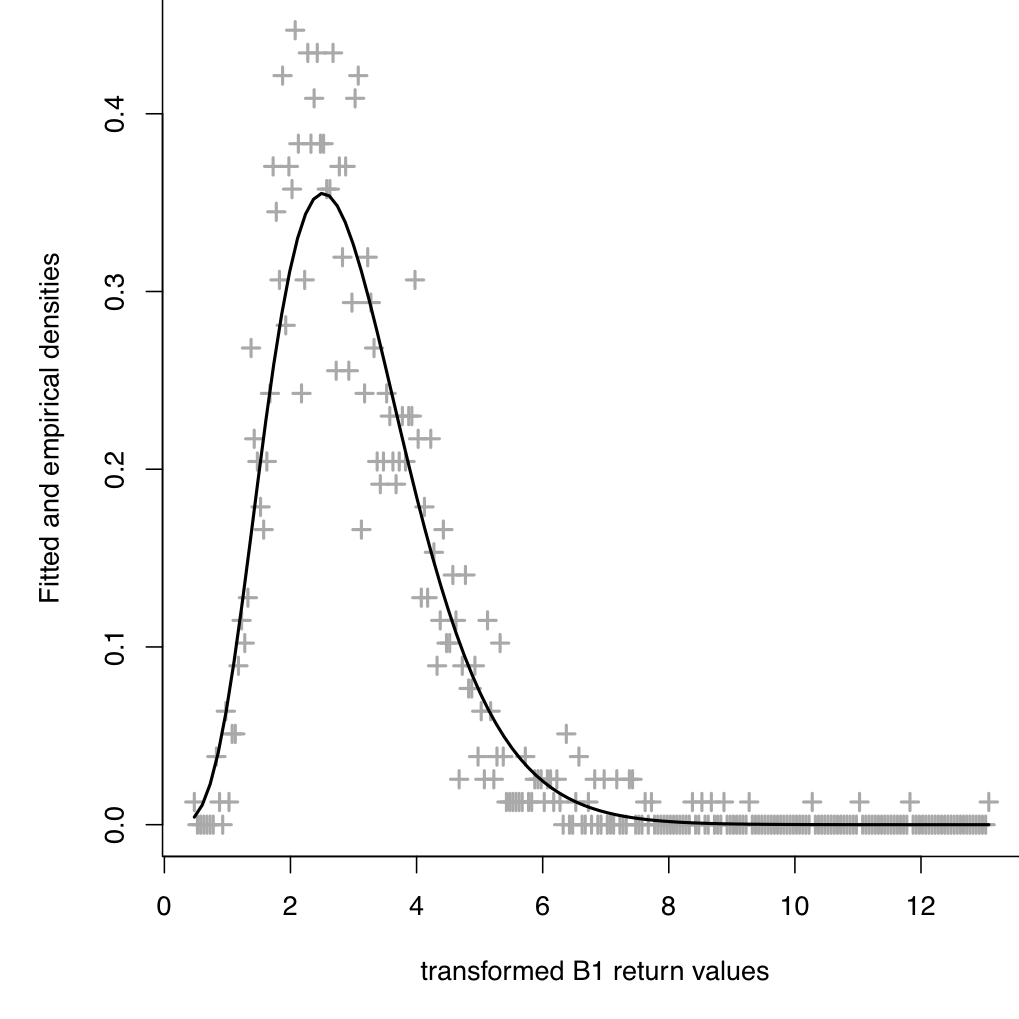}}
\subfigure[Region B2\label{la2}]
{\includegraphics[width=.48\linewidth]{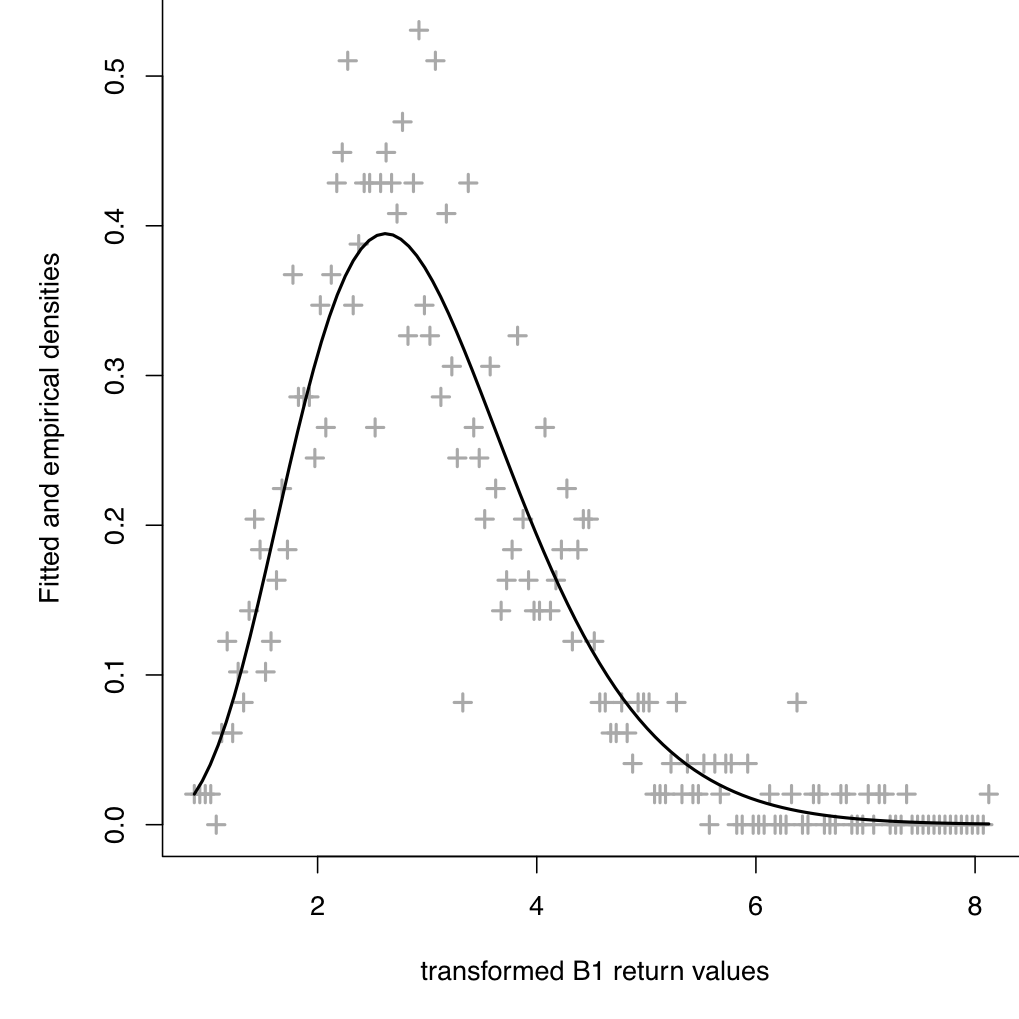}}
\subfigure[Region B3\label{la3}]
{\includegraphics[width=.48\linewidth]{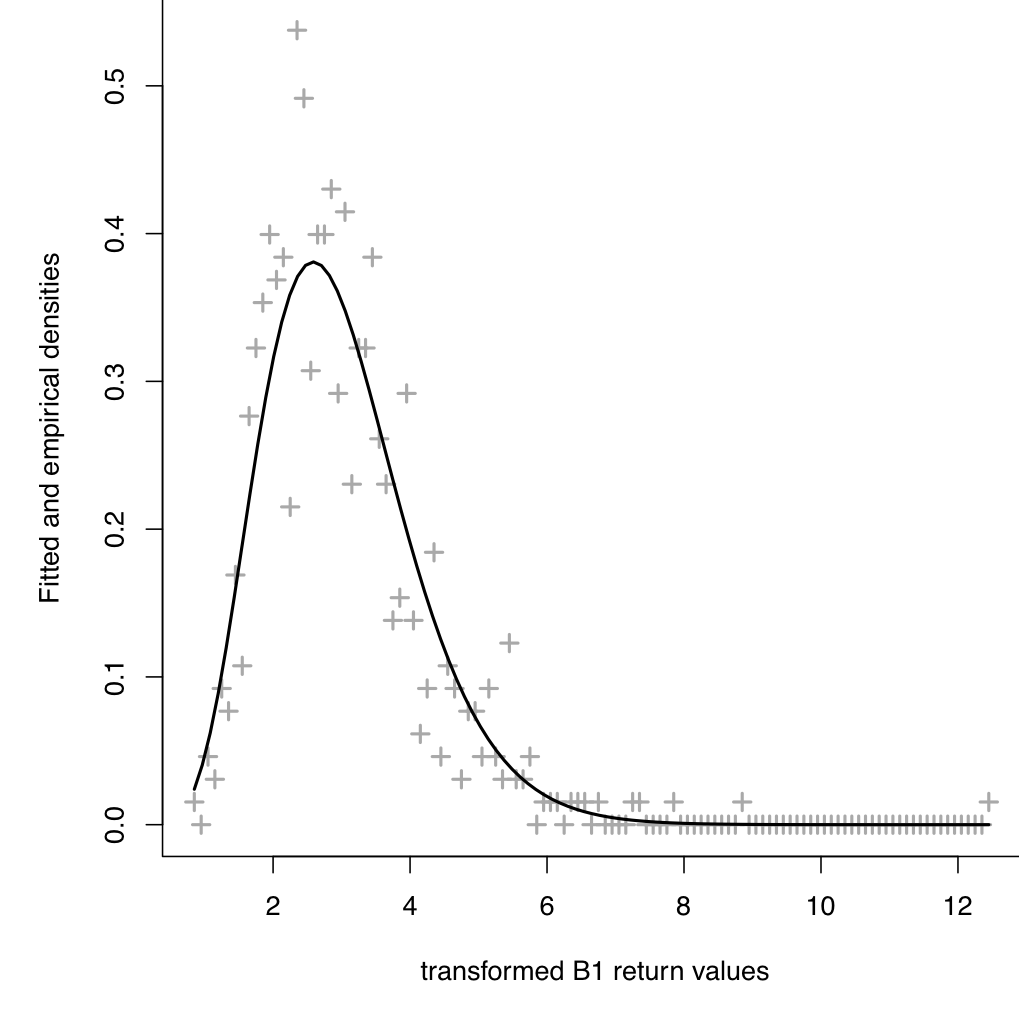}}
\caption{
Empirical (+) and fitted (solid) densities for transformed coherence matrices in selected regions.
}
\label{lat}
\end{figure}

According to Akbari~\emph{et~al.}~\cite{Akbarietal2013}, if $\{\bm{Z}_i;i=1,2,\ldots,n\}$ is a random sample drawn from $\bm{Z}\sim\mathcal{W}(L,\bm{\Sigma})$ and $\widehat{\bm{\Sigma}}$ represents the maximum likelihood estimator of $\bm{\Sigma}$, 
then $\operatorname{tr}(\widehat{\bm{\Sigma}}^{-1}\bm{Z}_i)$ follows a Gamma distribution for $i=1,2,\ldots,n$.
Fig.~\ref{lat} displays fitted and empirical densities of such transformed data for the three selected regions.
These results indicate that data may follow a scaled complex Wishart model.
Additionally, the Kolmogorov-Smirnov statistic $p$-values for checking the adequacy of the Gamma model to the transformed data
are $0.1377$, $0.4923$, and  $0.3911$ for regions B$_1$, B$_2$, and B$_3$, respectively.

We used likelihood ratio tests for two and three samples in order to quantify the similarity among these samples, 
The results presented in Table~\ref{discussion_regions} point out that $B_1$ is different from $B_2$ and $B_3$, but these last two are similar.
Although a visual inspection of areas~B$_2$ and~B$_3$ (Fig.~\ref{Apli10}) suggests regions of different nature, their observations projected via $\mathrm{tr}(\bm{\Sigma}^{-1}\bm{Z}_i)$ 
are statistically similar.

\begin{table}[hbt]
\centering   
\caption{Homogeneity test among considered samples}
\label{discussion_regions}
\begin{tabular}{crc}
\toprule
$\mathcal{H}_0$ & Statistics & $p$-value \\ 
\midrule
$B1=B2$ & $17.56$ & $1.53\times 10^{-5}$\\
$B1=B3$ & $5.90$ & $5.24\times 10^{-2}$ \\
$B2=B3$ & $1.31$ & $0.52$ \\
$B1=B2=B3$ & $19.00$ & $7.85\times 10^{-4}$ \\
\bottomrule
\end{tabular}
\end{table}

These samples are used to validate our proposed methods in Section~\ref{comparison:results}.

\section{Hypothesis tests in PolSAR Data: A Survey}\label{comparison:methodology}

This section provides a survey concerning three hypothesis tests which have been studied in the PolSAR data literature.

We assume that PolSAR data follow a scaled complex Wishart distribution.
Change detection is often formulated as a statistical test for
$\mathcal H_0: \boldsymbol{\Sigma}_1=\boldsymbol{\Sigma}_2$
assuming $L$ known.

The two main approaches in the literature are: 
(i)~likelihood ratio~\cite{Conradsen2003}
and 
(ii)~stochastic distances~\cite{FreryCintraNascimento2014}.
In this paper, the former proposal is extended to the context of scaled complex Wishart distributions, since the original approach used the nonscaled Wishart law.
Moreover, this paper also introduces an alternative way for validating $\mathcal H_0$ by means of entropy measures~\cite{FreryCintraNascimento2012}. 
Subsequently, these methodologies are introduced and discussed. 
In order to obtain more general results,
we will provide expressions for testing
$\mathcal H_0: 
(\boldsymbol{\Sigma}_1,L_1)
=
(\boldsymbol{\Sigma}_2,L_2)
$.

\subsection{Likelihood Ratio Statistics}

The log-likelihood ratio (LR) statistic has great importance in inference on parametric models. 
Let $S_{\text{LR}}$ be the LR statistic for assessing the simple null hypothesis $\mathcal H_0$.
As discussed in~\cite{caselaberge2002}, such statistic based on $\mathcal H_0$ has an asymptotic distribution $\chi^2_q$,
where $q$ is the difference between the dimensions of the parameter spaces under the alternative and the null hypotheses.
We denote such spaces by $\Theta_1$ and $\Theta_0$,
respectively.

Let $\{\boldsymbol{X}_1,\boldsymbol{X}_2,\ldots,\boldsymbol{X}_{N_1}\}$ and  $\{\boldsymbol{Y}_1,$ $\boldsymbol{Y}_2,$ $\ldots,$ $\boldsymbol{Y}_{N_2}\}$ be  two random samples from $\mathcal{W_R}(\boldsymbol{\Sigma}_1,L_1)$ and $\mathcal{W_R}(\boldsymbol{\Sigma}_2,L_2)$ of sizes $N_1$ and $N_2$, respectively.
The LR statistic is given by   
$$
S_{\text{LR}}=-2\log\lambda_{\mathcal{W_R}(\boldsymbol{\Sigma},L)},
$$
where 
$
\lambda_{\mathcal{W_R}(\boldsymbol{\Sigma},L)}
=
\sup_{\bm{\theta}\in\Theta_0} \ell(\bm{\theta})/\sup_{\bm{\theta}\in\Theta} \ell(\bm{\theta})$, 
$\Theta=\Theta_0\cup\Theta_1$,
and
$\Theta_0\cap\Theta_1=\varnothing$.
Thus,
we have that
\begin{align}
\log\,&\lambda_{\mathcal{W_R}(\boldsymbol{\Sigma},L)}
= 
\mathcal A(p) + \log\frac{|\widehat{\boldsymbol{\Sigma}}_1|^{N_1\,\widehat{L}_1}|\widehat{\boldsymbol{\Sigma}}_2|^{N_2\,\widehat{L}_2}}{ |\widehat{\boldsymbol{\Sigma}}_c|^{(N_1+N_2)\,\widehat{L}_c}}\nonumber \\
&+(\widehat{L}_c-\widehat{L}_1)\displaystyle \sum_{i=1}^{N_1}\log|\boldsymbol{X}_i| + (\widehat{L}_c-\widehat{L}_2)\displaystyle \sum_{i=1}^{N_2}\log|\boldsymbol{Y}_i|\nonumber \\
&+\displaystyle\sum_{i=1}^{N_1}\operatorname{tr}\bigl[(\widehat{L}_1\widehat{\boldsymbol{\Sigma}}_1^{-1}-\widehat{L}_c\widehat{\boldsymbol{\Sigma}}_c^{-1})\boldsymbol{X}_i\bigr]\nonumber \\
&+\displaystyle\sum_{i=1}^{N_2}\operatorname{tr}\bigl[(\widehat{L}_2\widehat{\boldsymbol{\Sigma}}_2^{-1}-\widehat{L}_c\widehat{\boldsymbol{\Sigma}}_c^{-1})\boldsymbol{Y}_i\bigr],
\label{SRex}
\end{align}
and
$$
\mathcal A(p)=p\log\frac{L_c^{L_c(N_1+N_2)}}{L_1^{L_1N_1}L_2^{L_2N_2}}+\log\frac{ \Gamma_p(L_1)^{N_1}\Gamma_p(L_2)^{N_2} }{ \Gamma_p(L_c)^{N_1+N_2} },
$$
where $L_c$ and $\boldsymbol{\Sigma}_c$ represent the number of looks and covariance matrix under the null hypothesis, respectively.
Akbari~\emph{et~al.}~\cite{Akbarietal20152}
discuss the two-sample LR test
under the $\mathcal{W_R}$ model.

Sections III-B and -C discuss tests for $\mathcal{H}_0$ based on
information-theoretic measures.

\subsection{The Kullback-Leibler distance}

The Kullback-Leibler divergence ($D_{\text{KL}}$) is one of oldest discrepancy measures between stochastic models; it has a central role in Information Theory~\cite{est3}.
This quantity was firstly understood as a measure of the error in choosing a model when another is the true one.
It has been used in image processing for 
segmentation~\cite{BeaulieuTouzi2004}, 
classification~\cite{KerstenandLeeandAinsworth2005}, 
boundary detection~\cite{PolarimetricSegmentationBSplinesMSSP,GambiniandMejailandJacobo-BerllesandFrery}, 
and change detection~\cite{IngladaMercier2007}.
Moreover, $D_{\text{KL}}$ has a close relationship with the Neyman-Pearson lemma~\cite{est3}, 
and its symmetrization has been suggested as a correction form 
for another important goodness-of-fit measure for comparing statistical models:
the Akaike information criterion~\cite{SeghouaneAmari2007}.

Let $\boldsymbol{X}$ and $\boldsymbol{Y}$ be two random matrices defined over the common support $\boldsymbol{\mathcal{X}}$ of positive definite complex matrices of size $p\times p$.
The Kullback-Leibler distance is defined by 
\begin{align*} 
d_{\text{KL}}(\boldsymbol{X},\boldsymbol{Y})&=\frac 12 [D_{\text{KL}}(\boldsymbol{X},\boldsymbol{Y})+D_{\text{KL}}(\boldsymbol{Y},\boldsymbol{X})]\\
&= \frac 12 \biggl[ \int_{\boldsymbol{\mathcal{X}}} f_{\boldsymbol{X}}\log{\frac{f_{\boldsymbol{X}}}{f_{\boldsymbol{Y}}}} \mathrm{d}\dot{\boldsymbol{Z}} + \int_{\boldsymbol{\mathcal{X}}} f_{\boldsymbol{Y}}\log{\frac{f_{\boldsymbol{Y}}}{f_{\boldsymbol{X}}}} \mathrm{d}\dot{\boldsymbol{Z}} \biggl] \\
&=\frac{1}{2}\int_{\boldsymbol{\mathcal{X}}}(f_{\boldsymbol{X}}-f_{\boldsymbol{Y}})\log{\frac{f_{\boldsymbol{X}}}{f_{\boldsymbol{Y}}}}\mathrm{d}\dot{\boldsymbol{Z}},
\end{align*}
with differential element $\mathrm{d}\dot{\boldsymbol{Z}}$ given by
$$
\mathrm{d}\dot{\boldsymbol{Z}}=\prod_{i=1}^p\mathrm{d}z_{ii}\prod^p_{\underbrace{i,j=1}_{i<j}}\mathrm{d}\Re\{z_{ij}\} \mathrm{d}\Im\{z_{ij}\},
$$ 
where $z_{ij}$ is the $(i,j)$-th entry of matrix $\dot{\boldsymbol{Z}}$; and $\Re$ and $\Im$ denote the real and imaginary part operators, respectively~\cite{Goodmanb}.

When distances are taken between particular cases of the same distribution, only the parameters are relevant.
In this case, the parameters $\boldsymbol{\theta_1}$ and $\boldsymbol{\theta_2}$ replace the random variables $\boldsymbol{X}$ and $\boldsymbol{Y}$.

Salicr\'u~\emph{et~al.}~\cite{salicruetal1994} proposed a hypothesis test based on $d_\text{KL}$.
Let $\widehat{\boldsymbol{\theta}}_1=(\widehat{\theta}_{11},\widehat{\theta}_{12},\ldots,\widehat{\theta}_{1M})^\top$ and $\widehat{\boldsymbol{\theta}}_2=(\widehat{\theta}_{21},\widehat{\theta}_{22},\ldots,\widehat{\theta}_{2M})^\top$ be the ML estimators for $\boldsymbol{\theta_1}$ and $\boldsymbol{\theta_2}$ based on random samples of size $N_1$ and $N_2$, respectively.
Under the regularity conditions discussed in \cite[p. 380]{salicruetal1994}, the following lemma holds.
\begin{Lemma}\label{prop-chi}
If $\frac{N_1}{N_1+N_2} \xrightarrow[N_1,N_2\rightarrow\infty]{} \lambda\in(0,1)$ and $\boldsymbol{\theta}_1=\boldsymbol{\theta}_2$, then
\begin{equation}
S_{\text{KL}}(\widehat{\boldsymbol{\theta}}_1,\widehat{\boldsymbol{\theta}}_2)=\frac{2 N_1 N_
Y}{N_1+N_2}
\frac{d_\text{KL}(\widehat{\boldsymbol{\theta}}_1,\widehat{\boldsymbol{\theta}}_2)}{ h{'}(0) \phi{''}(1)}   
\xrightarrow[N_1,N_2\rightarrow\infty]{\mathcal D}\chi_{M}^2,
\label{eq:chi2stat1}
\end{equation}
where ``$\xrightarrow[]{\mathcal{D}}$'' denotes convergence in distribution.
\end{Lemma}

Proposition~\ref{p-3} is a test for the null hypothesis $\boldsymbol{\theta}_1=\boldsymbol{\theta}_2$ based on Lemma~\ref{prop-chi}.

\begin{Proposition}
Let $S_{\text{KL}}(\widehat{\boldsymbol{\theta}}_1,\widehat{\boldsymbol{\theta}}_2)=s$ and $\widehat{\boldsymbol{\theta}}_1$ and $\widehat{\boldsymbol{\theta}}_2$ be ML estimates obtained from two sufficiently large random samples of sizes $N_1$ and $N_2$, respectively; then the null hypothesis $\boldsymbol{\theta}_1=\boldsymbol{\theta}_2$ can be re\-jec\-ted at le\-vel $\alpha$ if $\Pr( \chi^2_{M}>s)\leq \alpha$. 
\label{p-3}
\end{Proposition}

Frery~\emph{et~al.}~\cite{FreryCintraNascimento2014} presented closed expressions for $d_\text{KL}$ when
the random matrices $\boldsymbol{X}$ and $\boldsymbol{Y}$ follow
the Wishart distribution:
\begin{align}%
d_\text{KL}&(\boldsymbol{\theta}_1,\boldsymbol{\theta}_2)=\frac{L_1-L_2}{2}\bigg\{\log\frac{|\boldsymbol{\Sigma}_1|}{|\boldsymbol{\Sigma}_2|}-p\log\frac{L_1}{L_2}\nonumber \\
&+\psi_p^{(0)}(L_1)-\psi_p^{(0)}(L_2)\bigg\}-\frac{p(L_1+L_2)}{2}\nonumber \\
&+ \frac{\operatorname{tr}(L_2\boldsymbol{\Sigma}_2^{-1}\boldsymbol{\Sigma}_1+L_1\boldsymbol{\Sigma}_1^{-1}\boldsymbol{\Sigma}_2)}{2},
\label{expreKL}
\end{align}
from which the $S_{\text{KL}}$ test statistic follows.

\subsection{Shannon and R\'enyi Entropies}

The Shannon entropy has achieved a prominent position in PolSAR imagery.
Morio~\emph{et~al.}~\cite{MorioRefregierGoudailFernandezDupuis2009} applied it for extracting features from polarimetric targets, assuming the circular Gaussian distribution.
The Shannon entropy has also been used for classifying PolSAR textures~\cite{CloudePottier1997,Yanetal2010}.
In the subsequent discussion, we present a comprehensive examination of hypothesis tests based on Shannon and R\'enyi entropies. 

Let $f_{\boldsymbol{Z}}(\boldsymbol{Z};\boldsymbol{\theta})$ be a pdf with parameter vector $\boldsymbol{\theta}$.
The Shannon and R\'enyi (with order $\beta$) entropies are defined, respectively, as:
\begin{align}
H_{\text{S}}(\boldsymbol{\theta})=&-\int_{\boldsymbol{\mathcal{X}}} f_{\boldsymbol{Z}}(\dot{\boldsymbol{Z}};\boldsymbol{\Sigma},L)\log f_{\boldsymbol{Z}}(\dot{\boldsymbol{Z}};\boldsymbol{\Sigma},L) \,\mathrm{d}\dot{\boldsymbol{Z}}\nonumber \\
=&\operatorname{E}\{-\log f_{\boldsymbol{Z}}(\boldsymbol{Z})\}\label{entropy:1}\\
\text{and}\hskip+13ex&\nonumber \\
H_{\text{R}}^{\beta}(\boldsymbol{\theta})=&(1-\beta)^{-1}\log\int_{\boldsymbol{\mathcal{X}}} f_{\boldsymbol{Z}}^\beta(\dot{\boldsymbol{Z}};\boldsymbol{\Sigma},L)
 \mathrm{d}\dot{\boldsymbol{Z}}\nonumber \\
=&(1-\beta)^{-1}\log\operatorname{E}\bigl\{f_{\boldsymbol{Z}}^{\beta-1}(\boldsymbol{Z})\bigr\}\label{entropy:2}.
\end{align}

Pardo~\emph{et~al.}~\cite{Pardo1997} derived an important result which paves the way for asymptotic statistical inference methods based on entropies.
\begin{Lemma}\label{col1}
Let $\widehat{\boldsymbol{\theta}}=[\widehat{\theta_1}\;\widehat{\theta_2}\;\cdots\;\widehat{\theta_M}]^\top$ be the ML estimate of the parameter vector $\boldsymbol{\theta}=[\theta_1\;\theta_2\;\cdots\;\theta_M]^\top$ based on an $N$-point random sample from a model $\boldsymbol{Z}$ having pdf $f(\dot{\boldsymbol{Z}};\boldsymbol{\theta})$. 
Then
$$
\sqrt{N} \big[H_{\mathcal{M}}(\widehat{\boldsymbol{\theta}})-H_{\mathcal{M}}(\boldsymbol{\theta})\big] \xrightarrow[N\rightarrow \infty]{\mathcal D} \mathcal N(0,\sigma_{\mathcal{M}}^2(\boldsymbol{\theta})),
$$
where $\mathcal{M}\in\{\text{S},\text{R}\}$, $\mathcal N(\mu,\sigma^2)$ is the Gaussian distribution with mean $\mu$ and variance $\sigma^2$, 
\begin{equation}
\sigma_H^2(\boldsymbol{\theta})=\boldsymbol{\delta}^\top \mathcal K(\boldsymbol{\theta})^{-1}\boldsymbol{\delta},
\label{varentro}
\end{equation}
$\mathcal K(\boldsymbol{\theta})=\operatorname{E}\{-\partial^2 \log f_{\boldsymbol{Z}}(\boldsymbol{Z};\boldsymbol{\theta})/\partial \boldsymbol{\theta}^2\}$ is the Fisher information matrix, and $\boldsymbol{\delta}=[\delta_1\;\delta_2\;\cdots\;\delta_M]^\top$ such that 
$\delta_i=\partial H_{\mathcal{M}}(\boldsymbol{\theta})/\partial \theta_i$ for $i=1,2,\ldots,M$.
\end{Lemma}

Now we introduce a methodology for hypothesis tests and confidence intervals based on entropies.
We aim at testing the following hypotheses:
\begin{equation*}
\begin{cases}
\mathcal H_0 \colon
&
H_{\mathcal{M}}(\boldsymbol{\theta}_1)
=
H_{\mathcal{M}}(\boldsymbol{\theta}_2)=v
,
\\
\mathcal H_1 \colon
&
H_{\mathcal{M}}(\boldsymbol{\theta}_1)
\neq
H_{\mathcal{M}}(\boldsymbol{\theta}_2)
,
\end{cases}
\end{equation*}
where $\mathcal{M} \in \{\text{S},\text{R}\}$.
In other words, is there any statistical evidence for rejecting the assumption that two PolSAR samples come from the same model? 

Let $\widehat{\boldsymbol{\theta}_i}$ be the ML estimate for $\boldsymbol{\theta}_i$ based on a random sample of size $N_i$ from $\boldsymbol{Z}_i$ for $i=1,2,\ldots,r$ and $r\geq 2$.  
From Lemma~\ref{col1},
we have that 
$$
\sum_{i=1}^r\frac{N_i\big(H_{\mathcal M}(\widehat{\boldsymbol{\theta}_i})-\overline{v}\big)^2}{\sigma_{\mathcal M}^2(\widehat{\boldsymbol{\theta}_i})}\xrightarrow[N_i\rightarrow \infty]{\mathcal D} \chi^2_{r-1},
$$
where
$$
\overline{v}=\bigg[\sum_{i=1}^r\frac{N_i}{\sigma_{\mathcal M}^2(\widehat{\boldsymbol{\theta}_i})}\bigg]^{-1}\sum_{i=1}^r \frac{N_i\,H_{\mathcal M}(\widehat{\boldsymbol{\theta}_i})}{\sigma_{\mathcal M}^2(\widehat{\boldsymbol{\theta}_i})}.
$$
Then we obtain the following test statistic:
\begin{align}\label{teststatisticonentropy}
S_{\mathcal M}(\widehat{\boldsymbol{\theta}_1},\widehat{\boldsymbol{\theta}_2},\ldots,\widehat{\boldsymbol{\theta}_r})=\sum_{i=1}^r\frac{N_i\big(H_{\mathcal M}(\widehat{\boldsymbol{\theta}_i})-\overline{v}\big)^2}{\sigma_{\mathcal M}^2(\widehat{\boldsymbol{\theta}_i})};
\end{align} 
the expressions for 
$H_{\mathcal M}(\widehat{\boldsymbol{\theta}_i})$
and
$\sigma_{\mathcal M}^2(\widehat{\boldsymbol{\theta}_i})$
are presented the Appendix.
We are now in position to state the following result.

\begin{Proposition}\label{p1}
Let $N_i$, $i=1,2,\ldots,r$, be sufficiently large.
If $S_{\phi}^h(\widehat{\boldsymbol{\theta}_1},\widehat{\boldsymbol{\theta}_2},\ldots,\widehat{\boldsymbol{\theta}_r})=s$, then the null hypothesis $\mathcal H_0$ can be rejected at a level $\alpha$ if $\Pr\bigl( \chi^2_{r-1}>s\bigr)\leq \alpha$.
\end{Proposition}

Whereas tests based on stochastic distances, such as $d_{\text{KL}}$, allow contrasting only two samples, those based on entropies permit assessing $r$ samples at once; cf.~\eqref{teststatisticonentropy}.
For issues involving more than two populations ($r> 2$ in~\eqref{teststatisticonentropy}), this is a major advantage of the latter over the former.
In the case of comparing two samples of the same size, i.e. $r=2$ and $N_1=N_2=N$, \eqref{teststatisticonentropy} reduces to
$$
S_{\mathcal M}(\widehat{\boldsymbol{\theta}_1},\widehat{\boldsymbol{\theta}_2})
\,=\,
N\,
\frac{
[H_{\mathcal M}(\widehat{\boldsymbol{\theta}_1})\,-\,H_{\mathcal M}(\widehat{\boldsymbol{\theta}_2})]^2
}{
\sigma_{\mathcal M}^2(\widehat{\boldsymbol{\theta}_1})\,+\,\sigma_{\mathcal M}^2(\widehat{\boldsymbol{\theta}_2})
}.
$$

\section{Performance Analysis}\label{comparison:results}

In this section we assess the performance of the methodologies proposed with three experiments involving simulated (under the scaled Wishart complex law) and actual PolSAR data. 
Firstly, we use Monte Carlo experiments to measure 
(i)~test size (false alarm rate)
and
(ii)~test power ($1-\text{false negative rate}$).
For the test size,
we check whether two samples from $\bm{X}\sim W(\bm{B}_1,4)$
are from the same distribution, i.e., in a scenario where there was no change and there might be false positives.
We assess the test power
checking if
two samples from
$\bm{X}\sim W(\bm{B}_1,4)$
and
$\bm{X}\sim W(\bm{B}_1\cdot (1+k),4)$,
for $k=0.2,0.3,0.4$,
are correctly identified as a situation where there was a change.
We then perform two experiments with actual PolSAR data.

\subsection{Simulated Data}\label{comparison:resultsA}

We compare the following hypothesis tests:
\begin{itemize}
\item Likelihood ratio $S_{\text{LR}}$;
\item Kullback-Leibler distance $S_{\text{KL}}$;
\item Statistics based on Shannon $S_{\text{S}}$ and R\'enyi $S_{\text{R}}^\beta$ entropies.
\end{itemize}
We fixed $\beta=0.1$, since this value was found in Ref.~\cite{FreryCintraNascimento2012} to provide good discrimination in hard-to-deal-with situations.
We assume that the number of looks is known, as in~\cite{Conradsen2003,HypothesisTestingSpeckledDataStochasticDistances}.
Therefore, we are able to compare
information-theoretic measures with the methodology proposed by Conradsen~\emph{et~al.}~\cite{Conradsen2003}.

The samples are generated according to Algorithm~\ref{Algo:Wishart}.

\begin{algorithm}[hbt]
\caption{Sampling from the 	scaled complex Wishart distribution}\label{Algo:Wishart}
\begin{algorithmic}[1]
\Require $\bm{\Sigma}$ Hermitian positive definite $p\times p$ matrix  
\Require $L\geq 3$ integer
\State Denote  $\bm{R}=\Re\{\bm{\Sigma}\}$ and $\bm{I}=\Im\{\bm{\Sigma}\}$.
\For{$i=1,2,\ldots,L$}\label{first_step}
\State Generate an outcome of the $2p$-variate Gaussian distribution $\bm{x}_i=(x_{i1},x_{i2},\ldots,x_{ip},x_{i(p+1)},\ldots,x_{i(2p)})^\top\sim \mathcal N_{2p}(\bm{0},\bm{\Sigma^*})$, where 
\[ \bm{\Sigma^*} = \frac12 \Bigg[ \begin{array}{cc} \bm{R} & -\bm{I} \\ \bm{I} & \bm{R} \\ \end{array} \Bigg].\] 
\State Set the random vector 
$$
\bm{y}_i = (x_{i1},x_{i2},\ldots,x_{ip})^\top + \bm{j}\,\,(x_{i(p+1)},\ldots,x_{i(2p)})^\top.
$$ 
With this, $\bm{y}_i$ is a $p$-variate outcome of the complex Gaussian distribution $\mathcal{N}^{C}_p(\bm 0,\boldsymbol{\Sigma})$.
\EndFor
\State\label{StepReturn} Return
$ 
L^{-1} \sum_{i=1}^L \bm{y}_i \bm{y}_i^{*}
$,
outcome of ${\mathcal W}(\bm{\Sigma},L)$, the scaled complex Wishart distribution.
\end{algorithmic}
\end{algorithm}

The parameters used for assessing the null hypothesis $\mathcal{H}_0\colon \boldsymbol{\Sigma}_1=\boldsymbol{\Sigma}_2$ are
$L_1=L_2=4$, and~\eqref{samplecov}, the sample covariance matrix of area B$_1$, Fig.~\ref{Apli10}.
As we are interested in the behavior of the tests with small sample sizes, we computed the size of the hypothesis at $\alpha\in\{\SI{1}{\percent}, \SI{5}{\percent}, \SI{10}{\percent}\}$ for $N_1=N_2=N \in\{10,11,\ldots,50\}$.

Let $T$ be the number of Monte Carlo replications and $C$ the number of occurrences under $\mathcal H_0$ (i.e., pairs of samples are taken from the same model) on which the null hypothesis is rejected at the nominal level $\alpha$. 
The empirical test size (ETS) or false positive rate is defined by 
$
\alpha_{\text{ETS}}={C}/{T}
$.
We used 
$T=5500$, 
as suggested in~\cite{HypothesisTestingSpeckledDataStochasticDistances}, and $\alpha_{\text{ETS}}$ did not suffer expressive changes for larger values.

\begin{figure*}[hbt]
\begin{equation}
B_1={\left[\begin{array}{ccc} 9.528 \times 10^{-3} & (-3.469+1.048\,\textbf{j})\times 10^{-4} & (1.439+1.164\,\textbf{j})\times 10^{-3} \\ 
& 1.794 \times 10^{-3} & (8.551-1.608\,\textbf{j})\times 10^{-5} \\
&  & 4.955 \times 10^{-3}  \end{array} \right]}.
\label{samplecov}
\end{equation}
\hrulefill
\end{figure*}

Table~\ref{comparison:table1} shows: 
(i)~the empirical test size at nominal levels \SI{1}{\percent}, \SI{5}{\percent}, \SI{10}{\percent}, and
(ii)~the mean test statistic ($\overline{S}_{\bullet}$)
of the four statistics. 

\begin{table}[hbt]
\centering
\caption{Estimated Test Sizes (False Positive rates)}\label{comparison:table1}
\begin{tabular}{c rrr r} \toprule 
\multirow{2}{*}{$N$} & \multicolumn{4}{c}{Mean values} \\  \cmidrule(lr{.25em}){2-5}
  & \SI{1}{\percent} & \SI{5}{\percent} & \SI{10}{\percent} & $\overline{S}_{\bullet}$  \\                                        
\cmidrule(lr{.25em}){1-1} \cmidrule(lr{.25em}){2-4} \cmidrule(lr{.25em}){5-5} 
&  \multicolumn{4}{c}{Likelihood ratio ($S_{\text{LR}}$)}\\
$10-20$ & $1.21$ & $5.76$ & $11.16$ & $9.25$  \\
$21-30$ &  $1.10$ & $5.42$ & $10.66$ & $9.12$  \\
$31-40$ &  $1.03$ & $5.13$ & $10.43$ & $9.09$  \\
$41-50$ &  $1.06$ & $5.21$ & $10.28$ & $9.08$  \\ 
\rule{0pt}{1ex} &  \multicolumn{4}{c}{Shannon Entropy ($S_{\text{S}}$) }\\                                                
\rule{0pt}{.5ex} &  $1.00$ & $4.59$ & $9.47$  & $1.00$  \\
&  $0.99$ & $4.64$ & $9.21$  & $1.02$  \\
&  $0.99$ & $4.44$ & $9.24$  & $1.05$  \\
&  $0.93$ & $4.54$ & $9.37$  & $1.07$  \\
\rule{0pt}{1ex} & \multicolumn{4}{c}{R\'enyi Entropy ($S_{\text{R}}^{0.1}$) }\\                                          
\rule{0pt}{.5ex} &  $0.00$ & $1.71$ & $4.53$  & $0.71$  \\
&  $0.33$ & $1.83$ & $4.53$  & $0.74$  \\
&  $0.32$ & $1.79$ & $4.46$  & $0.77$  \\
&  $0.29$ & $1.78$ & $4.56$  & $0.79$  \\
\rule{0pt}{1ex} & \multicolumn{4}{c}{Kullback-Leibler Distance ($S_{\text{KL}}$)}\\                                       
\rule{0pt}{.5ex} &  $1.83$ & $7.06$ & $12.89$ & $9.53$  \\
&  $1.43$ & $6.16$ & $11.66$ & $9.27$  \\
&  $1.24$ & $5.67$ & $11.17$ & $9.20$  \\
&  $1.24$ & $5.55$ & $10.85$ & $9.16$  \\
\bottomrule
\end{tabular}
\end{table}

In average, all test statistics behave as expected when the sample sizes increase: $\bar{S}_{\text{S}}$ and $\bar{S}_{\text{R}}^{0.1}$ tend to one, while $\bar{S}_{\text{LR}}$ and $\bar{S}_{\text{KL}}$ tend to nine.
Recall that the asymptotic distribution of the two former is $\chi^2_9$, while the two latter are $\chi^2_1$.

\begin{figure*}[hbt]
\centering
\subfigure[\SI{1}{\percent}\label{Apli31}]{\includegraphics[width=.31\linewidth]{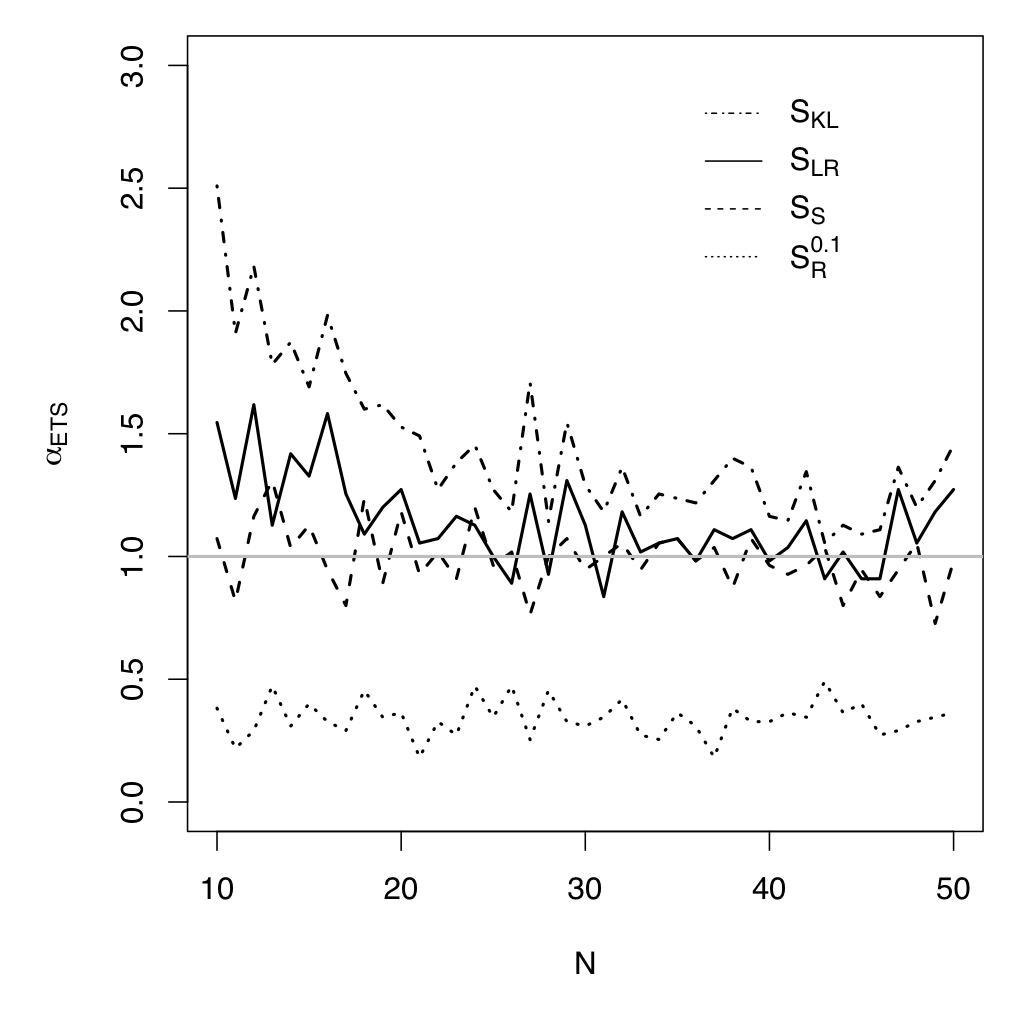}}
\subfigure[\SI{5}{\percent}, \label{Apli32}]{\includegraphics[width=.31\linewidth]{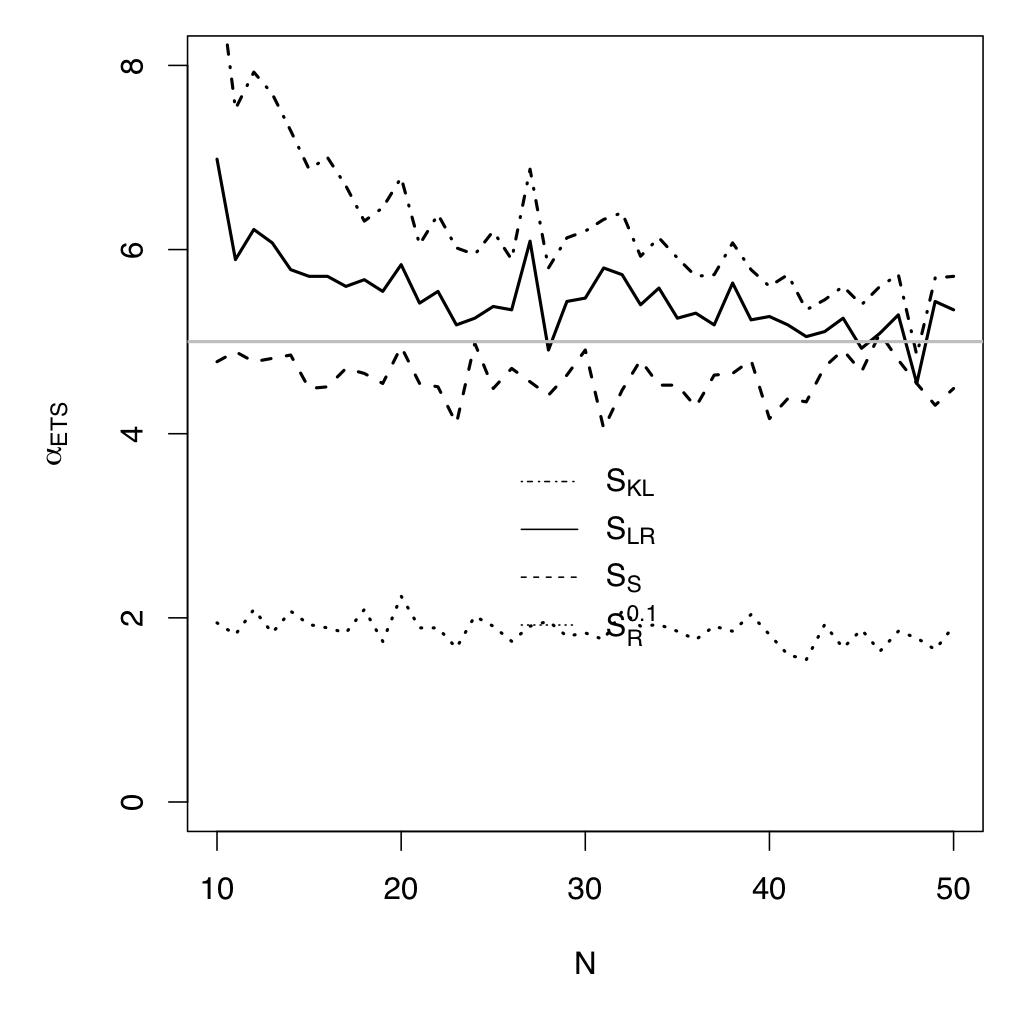}}
\subfigure[ \SI{10}{\percent}\label{Apli33}]{\includegraphics[width=.31\linewidth]{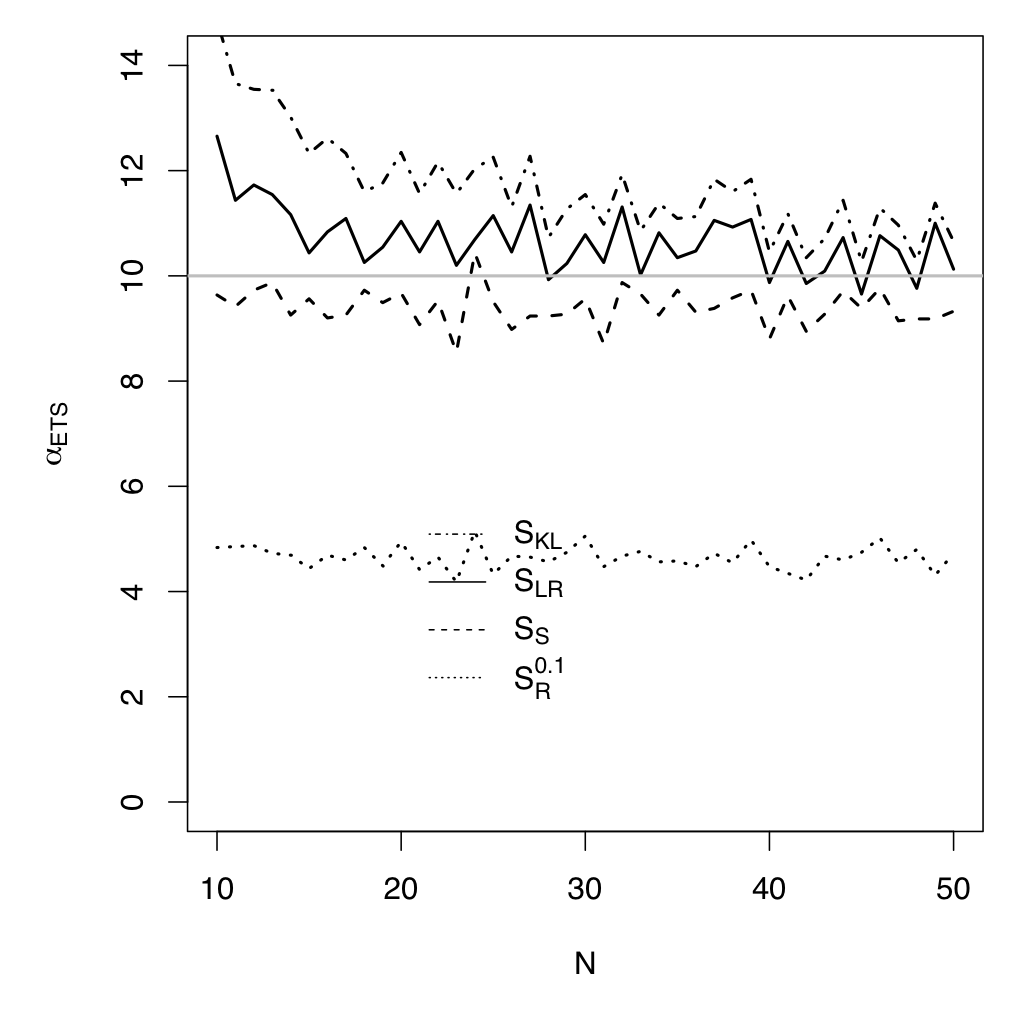}}
\caption{Values for $\alpha_{\text{ETS}}$ sizes on synthetic data for different scenarios at the levels \SI{1}{\percent}, \SI{5}{\percent}, \SI{10}{\percent}.} 
\label{Apli3}  
\end{figure*}

The $S_{\text{LR}}$ and $S_{\text{S}}$ tests exhibit the closest empirical sizes to the nominal levels, 
as confirmed by Fig.~\ref{Apli3}.
The ETS associated with $S_{\text{R}}^{0.1}$ and $S_{\text{KL}}$ are biased, however the bias reduces as the sample size increases.
We conclude that these two statistics require larger sample sizes to achieve the expected asymptotic behavior.

In general terms, Table~\ref{comparison:table1} suggests this inequality:
\begin{equation}
{\text{ETS}}_{S_{\text{KL}}}\geq
{\text{ETS}}_{S_{\text{LR}}}\geq
{\text{ETS}}_{S_{\text{S}}}\geq
{\text{ETS}}_{S_{\text{R}}^{0.1}}.
\label{TAM}
\end{equation}
The size of tests (False Positive rates) based on the Shannon entropy and likelihood ratio are the closest to the nominal level.



We also studied the test power.
We wish to reject the hypothesis $\mathcal H_0$ given two samples drawn from $\boldsymbol{X}\sim \mathcal W(B_1,4)$ and $\boldsymbol{Y}\sim \mathcal W(B_1\cdot (1+k),4)$ where $k=0.2,0.3,0.4$; i.e, 
under $\mathcal H_1$.
The rate
$
\eta={(T-C^*)}/T
$,
where $C^*$ is the number of rejections of $\mathcal H_0$ under
$\mathcal H_1$,
estimates the Type~II error or false negative~\cite{Conradsen2016}, and we aim at quantifying the test power $1-\eta$.

\begin{figure*}[hbt]
\centering
\subfigure[$(B_1,4)\text{ vs. }(B_1\cdot(1+0.2),4)$\label{Apli21}]{\includegraphics[width=.31\linewidth]{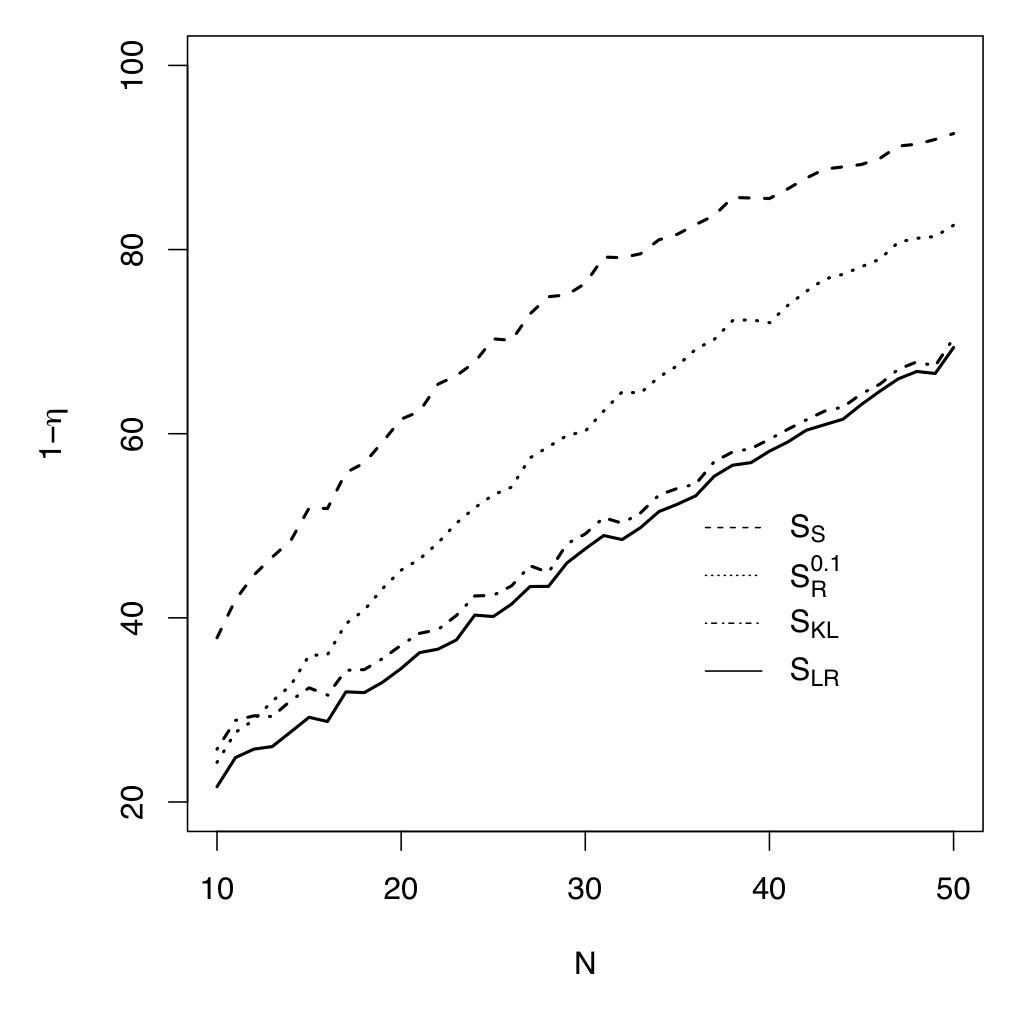}}
\subfigure[$(B_1,4)\text{ vs. }(B_1\cdot(1+0.3),4)$\label{Apli22}]{\includegraphics[width=.31\linewidth]{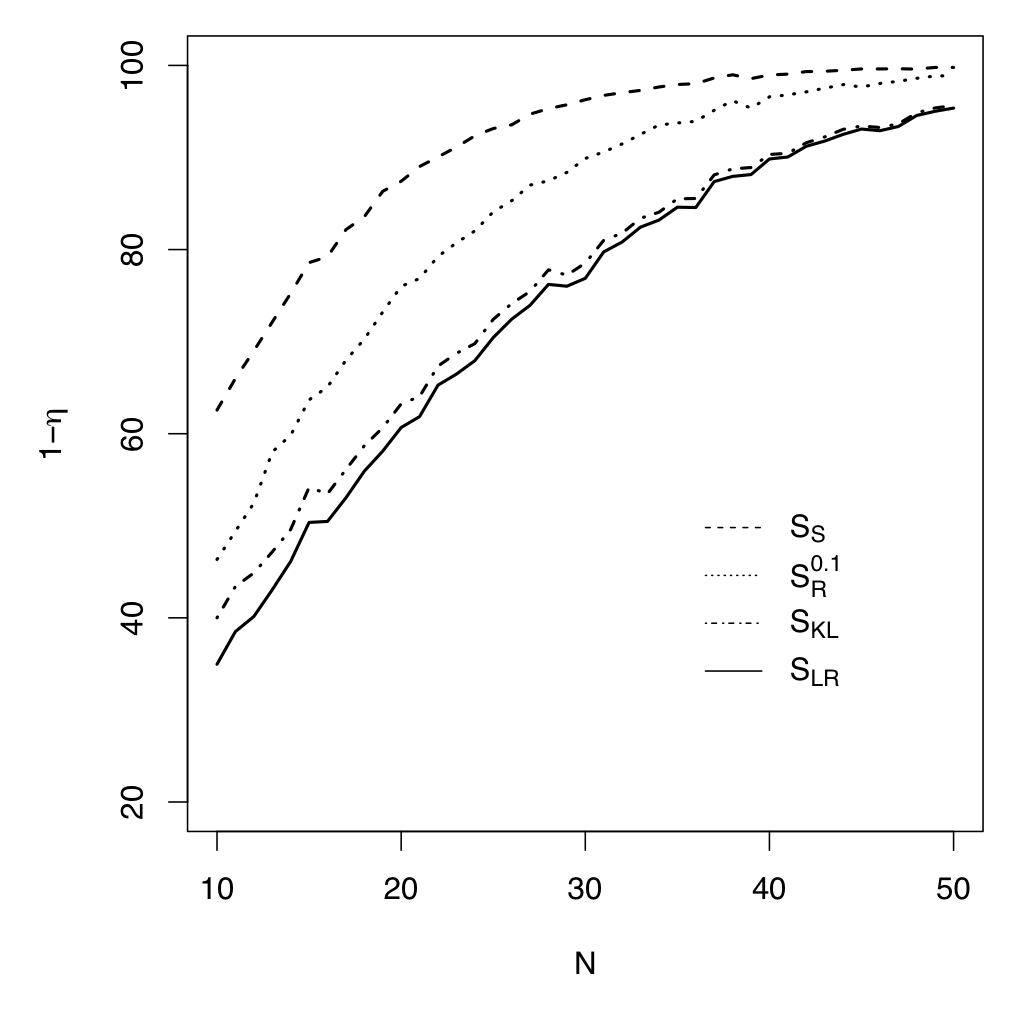}}
\subfigure[$(B_1,4)\text{ vs. }(B_1\cdot(1+0.4),4)$\label{Apli23}]{\includegraphics[width=.31\linewidth]{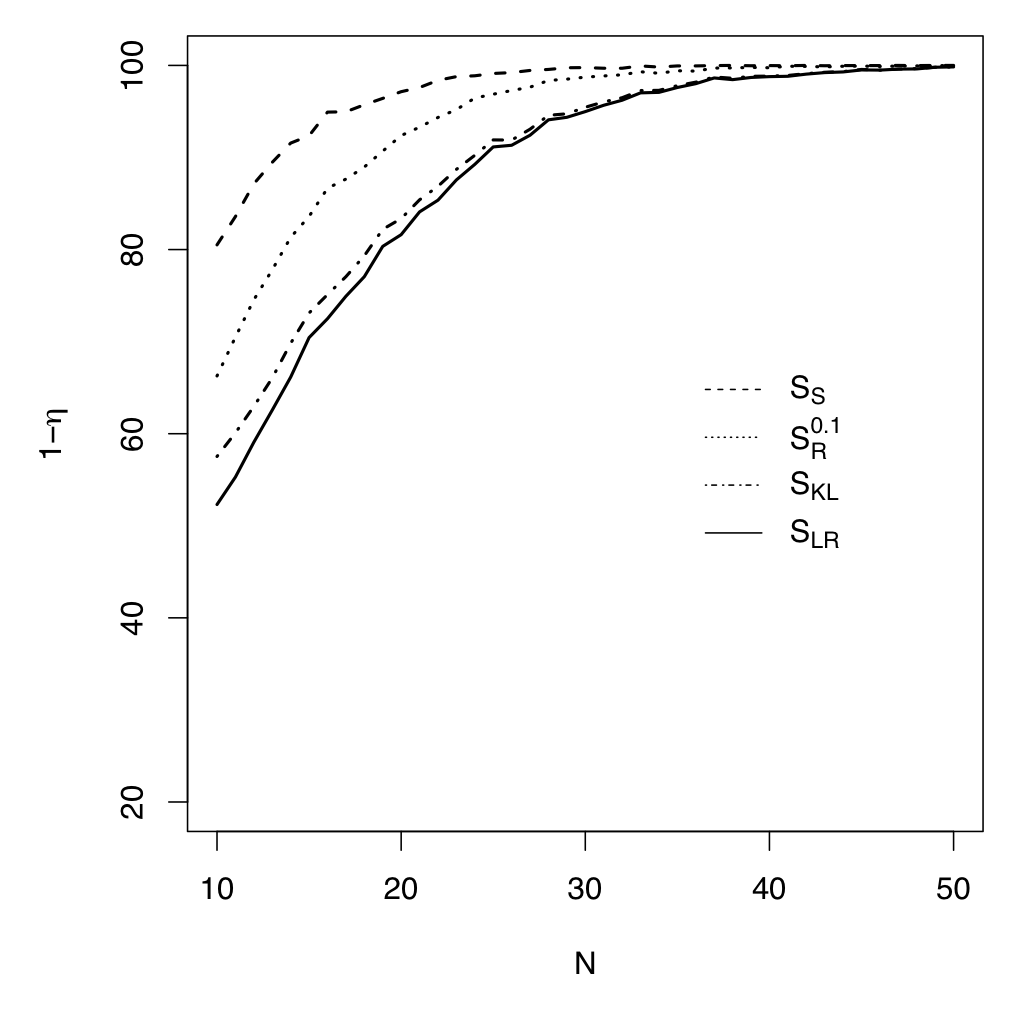}}
\caption{Estimated test powers for several scenarios at the level \SI{1}{\percent}.} 
\label{Apli2}  
\end{figure*}

Fig.~\ref{Apli2} presents the estimated power for several samples sizes. 
The test based on Shannon entropy performs best.
%
In this case, we obtain the inequality:
\begin{equation*}
({1-\eta})_{S_{\text{S}}}\geq
({1-\eta})_{S_{\text{R}}^{0.1}}\geq
({1-\eta})_{S_{\text{KL}}}\geq
({1-\eta})_{S_{\text{LR}}}.
\end{equation*}
The relation between discriminatory powers within groups
$\{S_{\text{KL}}, S_{\text{LR}}\}$ and $\{S_{\text{S}}, S_{\text{R}}^\beta\}$
has been discussed in the statistical literature.
This fact can be explained twofold, namely
(i)~the relationship between the Neyman and Pearson lemma and the Kullback-Leibler distance~\cite{est3}, and 
(ii)~the fact that 
$\lim_{\beta \to 1} S_{\text{R}}^\beta = S_{\text{S}}$~\cite{coverandthomas1991}.

The best test statistics should have both empirical size near to the nominal level, and the highest estimated power.
Thus, based on this evidence and on the estimated size, we suggest $S_{\text{S}}$ as the best discriminator on scenarios which follow the scaled complex Wishart distribution.

%
%
%

\subsection{Experiments with Data from Sensors}

In this section
we apply the proposed test statistics
to two studies:
(i)~for assessing $\mathcal H_0:\boldsymbol{\Sigma}_1=\boldsymbol{\Sigma}_2$ to the data presented in Fig.~\ref{Apli10} (single date)
and
(ii)~for detecting changes on two PolSAR images captured at different instants, as displayed in Fig.~\ref{Application2} (multitemporal data).
ENL is assumed constant.

\subsubsection{Single date experiment}

Our first experiment aims at assessing $\alpha_{\text{ETS}}$, the empirical test size (Type~I error or Probability of False Alarm), using pairs of disjoint samples from the same target.

This experiment is outlined in Algorithm~\ref{applicationsteps}.
We used samples of size $N\in\{3\times 3, 4\times 4,\ldots,23\times 23\}$.

\begin{algorithm}[hbt]
\caption{Experiment design for data from the same target}\label{applicationsteps}
\begin{algorithmic}[1]
\For{$j=1,2,\ldots,5500$}
	\State\label{enui} Extract two disjoint regions $\boldsymbol{U}_j$ and $\boldsymbol{V}_j$ from areas  $\text{B}_1,\text{B}_2,\text{ and }\text{B}_3$.
	\State Generate two vectors of size $N$, $\boldsymbol{u}^{(j)}$ and $\boldsymbol{v}^{(j)}$ from $\boldsymbol{U}_j$ and $\boldsymbol{V}_j$, respectively, sampling without replacement.
	\State Estimate $\widehat{\boldsymbol{\theta}}^{(j)}_1$ and $\widehat{\boldsymbol{\theta}}^{(j)}_2$ based on $\boldsymbol{u}^{(j)}$ and $\boldsymbol{v}^{(j)}$, respectively.
	\State\label{enuf} Compute the decision from Propositions~\ref{p-3} and~\ref{p1}, and execute the test based on $S_{\text{LR}}$ for $\alpha =\{\SI{1}{\percent}, \SI{5}{\percent}, \SI{10}{\percent}\}$.
\EndFor
\State\label{last_step} Let $T$ be the number of times that the null hypothesis is rejected. 
Calculate the empirical test size ($\widehat{\alpha}_{1-\alpha}$) at level $\alpha$ as
$$
\alpha_{\text{ETS}}=T/5500,\text{ if }\boldsymbol{V}_j=\boldsymbol{U}_j.
$$
\end{algorithmic}
\end{algorithm}

\begin{figure}[hbt]
\centering
\subfigure[\SI{1}{\percent}\label{Aplii31}]{\includegraphics[width=.5\linewidth]{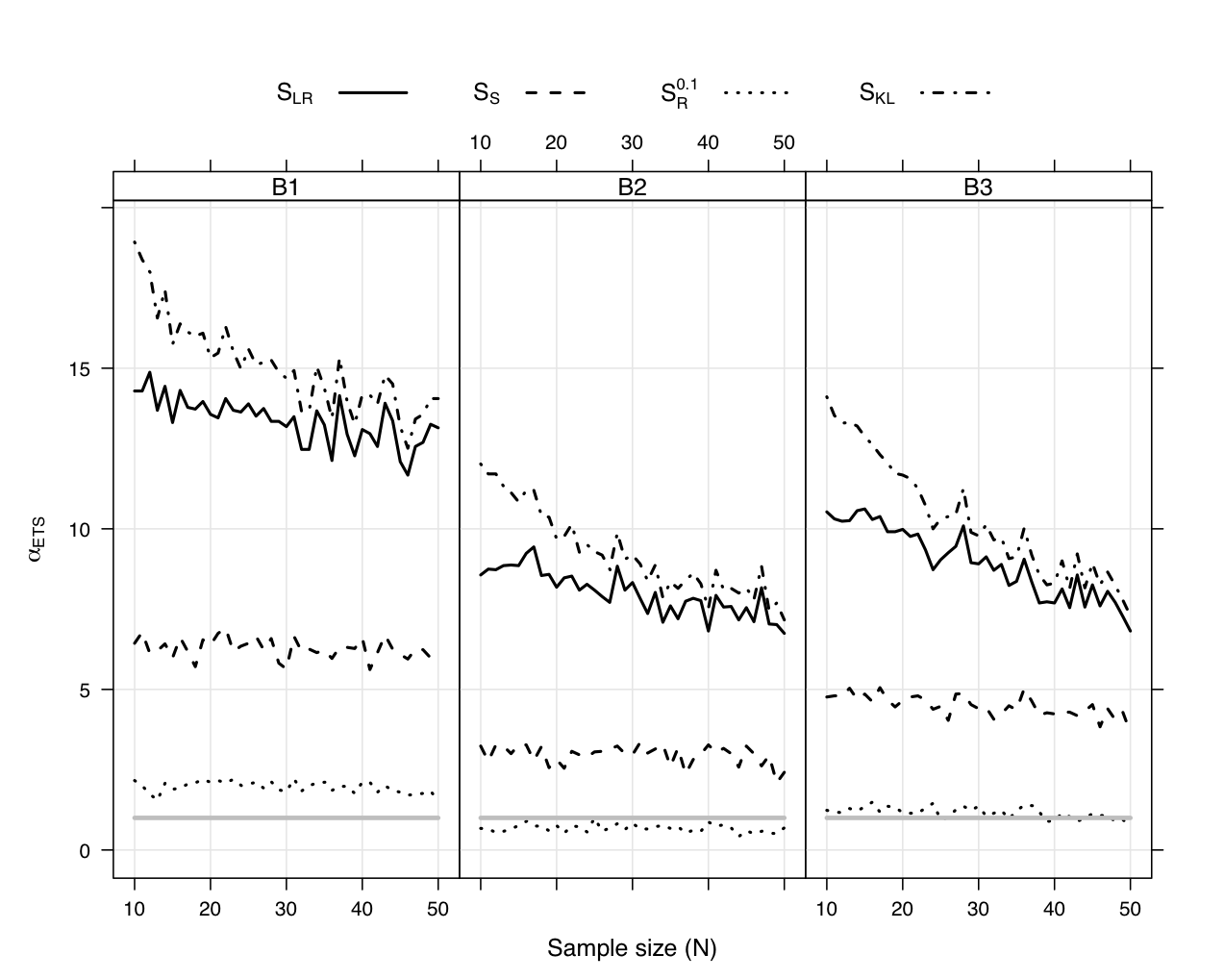}}
\subfigure[\SI{5}{\percent} \label{Aplii32}]{\includegraphics[width=.5\linewidth]{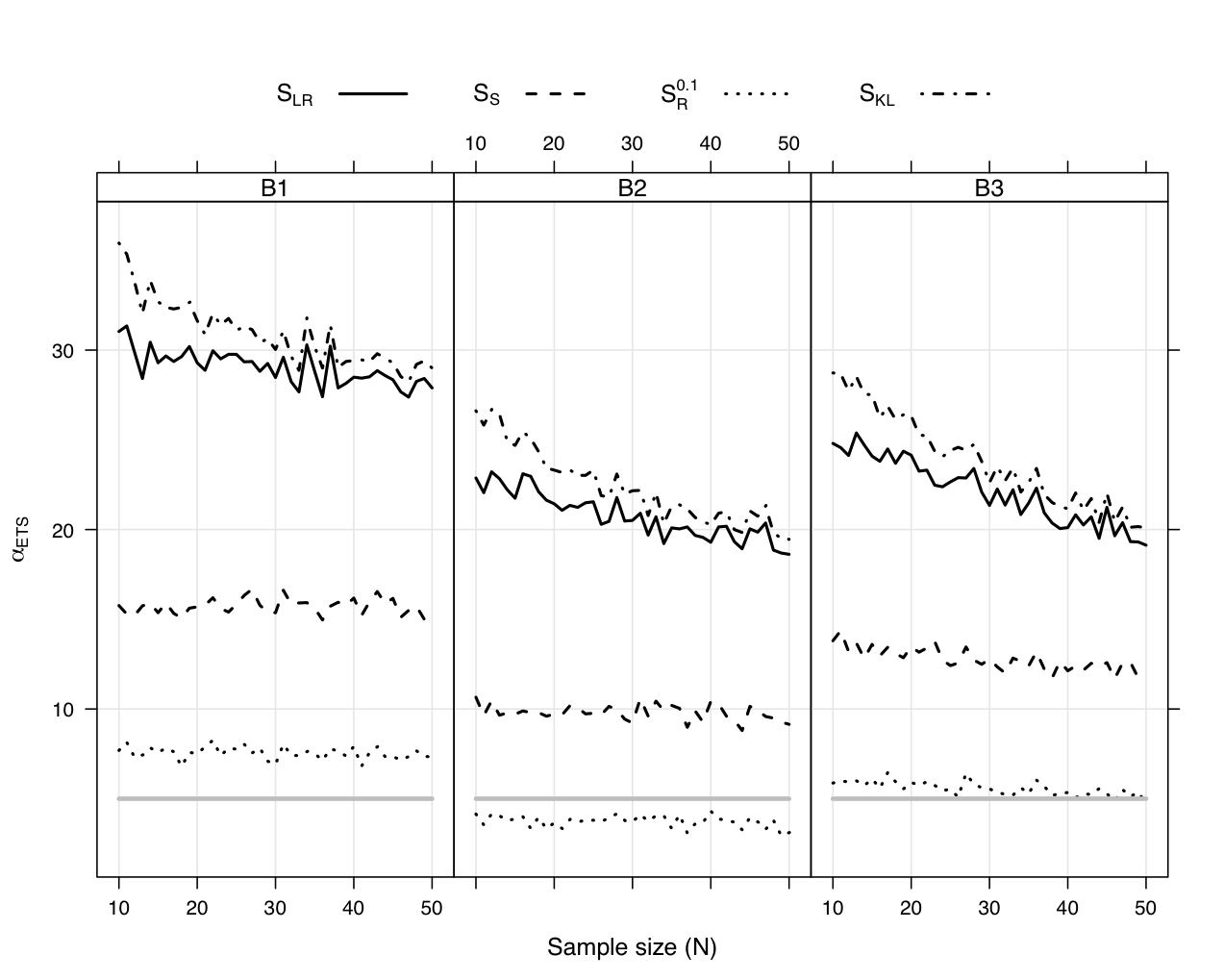}}
\subfigure[\SI{10}{\percent}\label{Aplii33}]{\includegraphics[width=.5\linewidth]{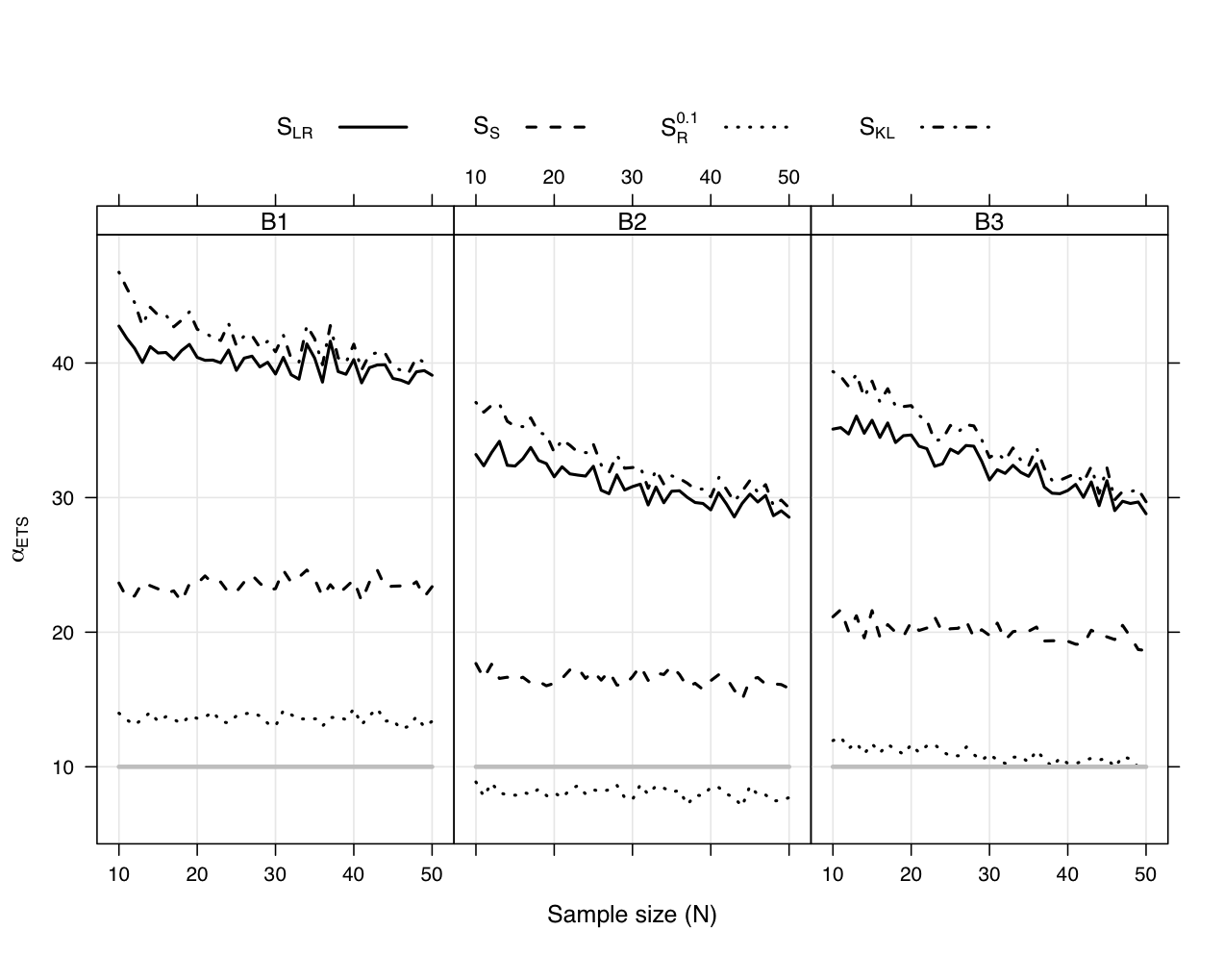}}
\caption{Empirical test size for actual data at levels \SI{1}{\percent}, \SI{5}{\percent}, \SI{10}{\percent}.}
\label{comparison2}
\end{figure}

Fig.~\ref{comparison2} shows the observed $\alpha_{\text{ETS}}$.
Inequality~\eqref{TAM} is also verified on actual data.
For $\alpha=\SI{1}{\percent}$,  
$\SI{11.67}{\percent}\leq\alpha_{\text{ETS}}(S_{\text{LR}})<\alpha_{\text{ETS}}(S_{\text{KL}})\leq \SI{18.93}{\percent}$
and
$\SI{1.527}{\percent}\leq\alpha_{\text{ETS}}(S_{\text{S}})<\alpha_{\text{ETS}}(S_{\text{R}}^{0.1})\leq \SI{6.909}{\percent}$;
i.e., all tests overestimate $\alpha$,
but $S_{\text{S}}$ and $S_{\text{R}}^{0.1}$ presented better results than $S_{\text{LR}}$ and $S_{\text{KL}}$.
For B$_2$ and B$_3$, 
$\alpha_{\text{ETS}}(S_{\text{R}}^{0.1})\leq \SI{1.49}{\percent} $,
overcoming
$\SI{2.091}{\percent}\leq\alpha_{\text{ETS}}(S_{\text{S}})\leq \SI{5.055}{\percent}$, 
$\SI{6.745}{\percent}\leq\alpha_{\text{ETS}}(S_{\text{LR}})\leq \SI{10.618}{\percent}$
and  
$\SI{7.164}{\percent}\leq\alpha_{\text{ETS}}(S_{\text{KL}})\leq \SI{14.109}{\percent}$.

PolSAR regions
are Wishart, our explanation for the better performance of $S_{\text{R}}^{0.1}$ is deviations from this hypothesis.

These results present evidence that the test statistics based on $S_{\text{R}}^{0.1}$ outperforms the other ones.
This test presented good results even for small samples.
Thus, this measure is suggested as a relevant change detection tool for PolSAR imagery.

\subsubsection{Multitemporal data}

\begin{figure}[hbt]
\centering
\subfigure[First scene\label{fig:StudyArea:1}]{
\includegraphics[width=.45\linewidth]{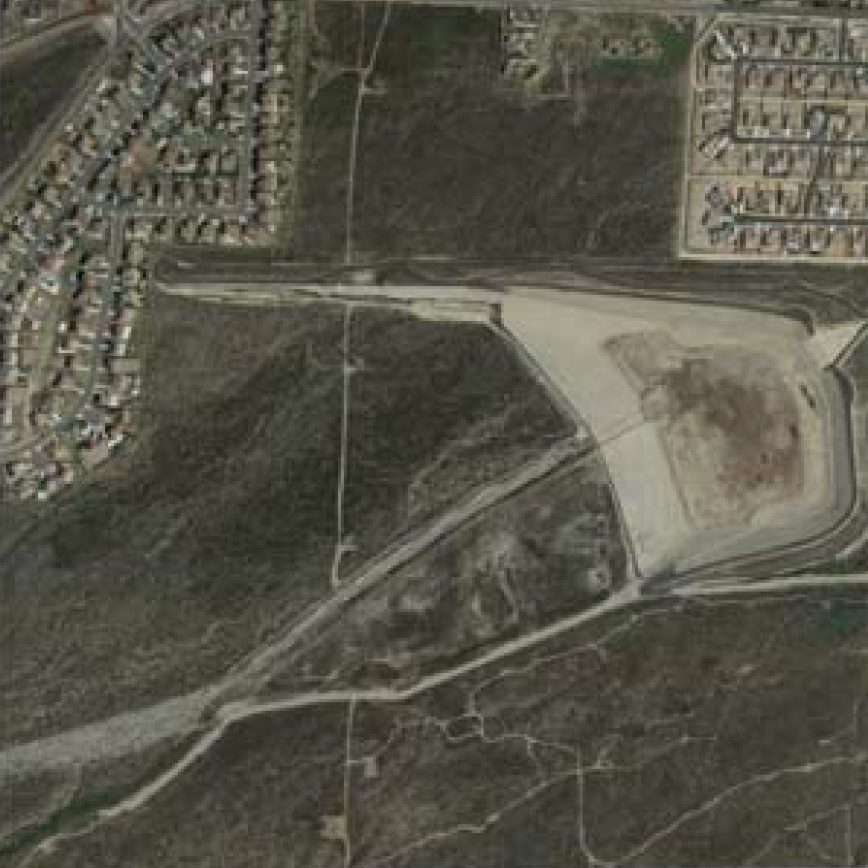}
}
\subfigure[Second scene\label{fig:StudyArea:2}]{
\includegraphics[width=.45\linewidth]{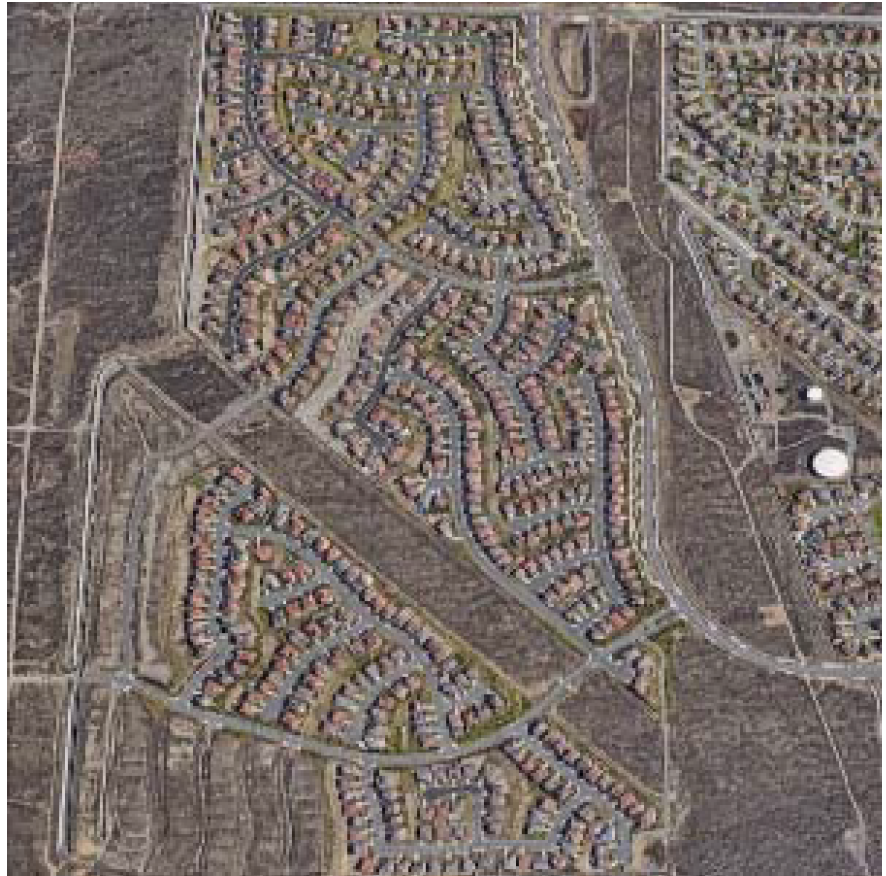}
}
\caption{
Images from the study areas: Los Angeles, California.
}
\label{fig:StudyArea}
\end{figure}

\begin{figure}[hbt]
\centering
\subfigure[Scene 1 (before)\label{Application2:1}]{
\includegraphics[width=.43\linewidth]{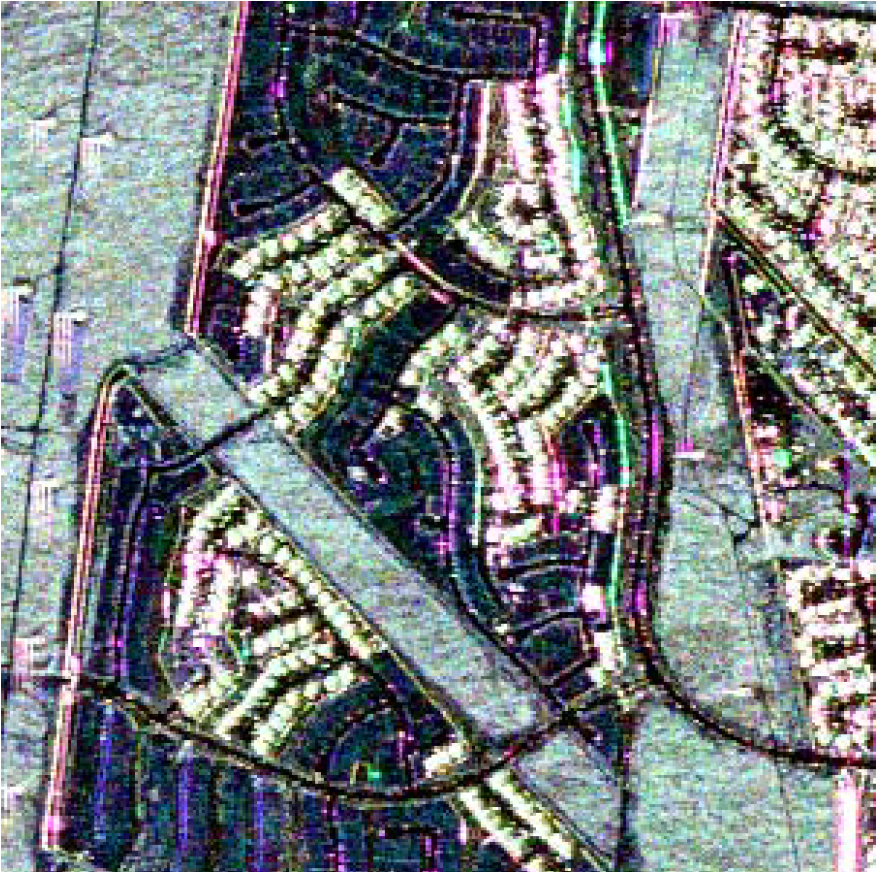}}
\subfigure[Scene 1 (after)\label{Application2:2}]{
\includegraphics[width=.43\linewidth]{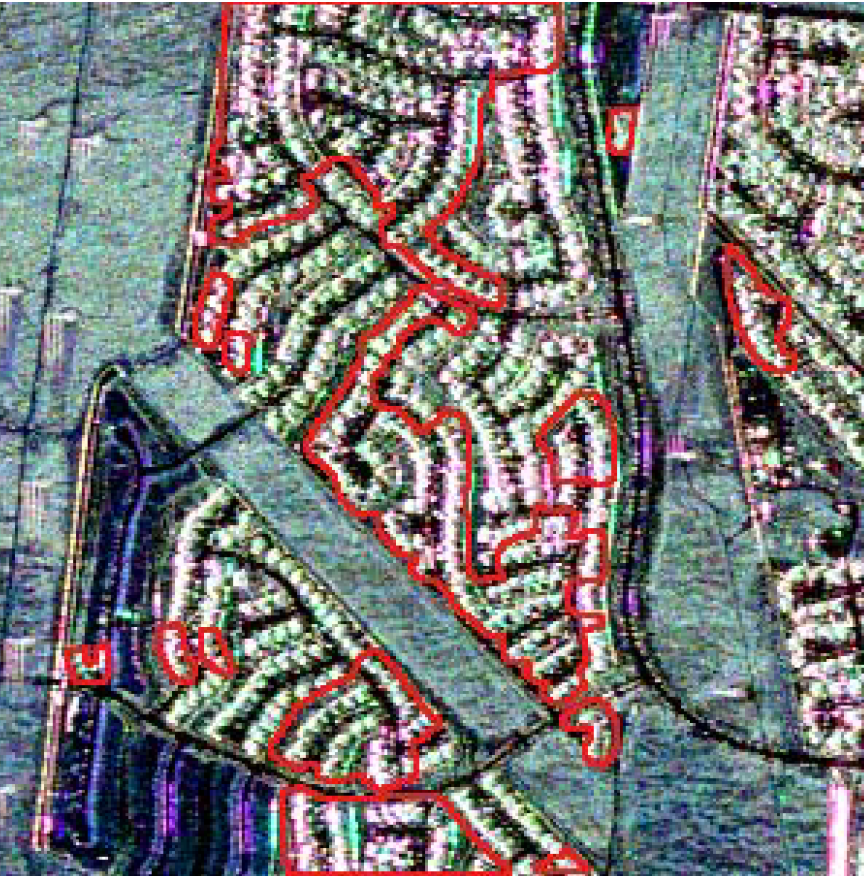}}
\\
\subfigure[Scene 2 (before)\label{Application2:3}]{
\includegraphics[width=.43\linewidth]{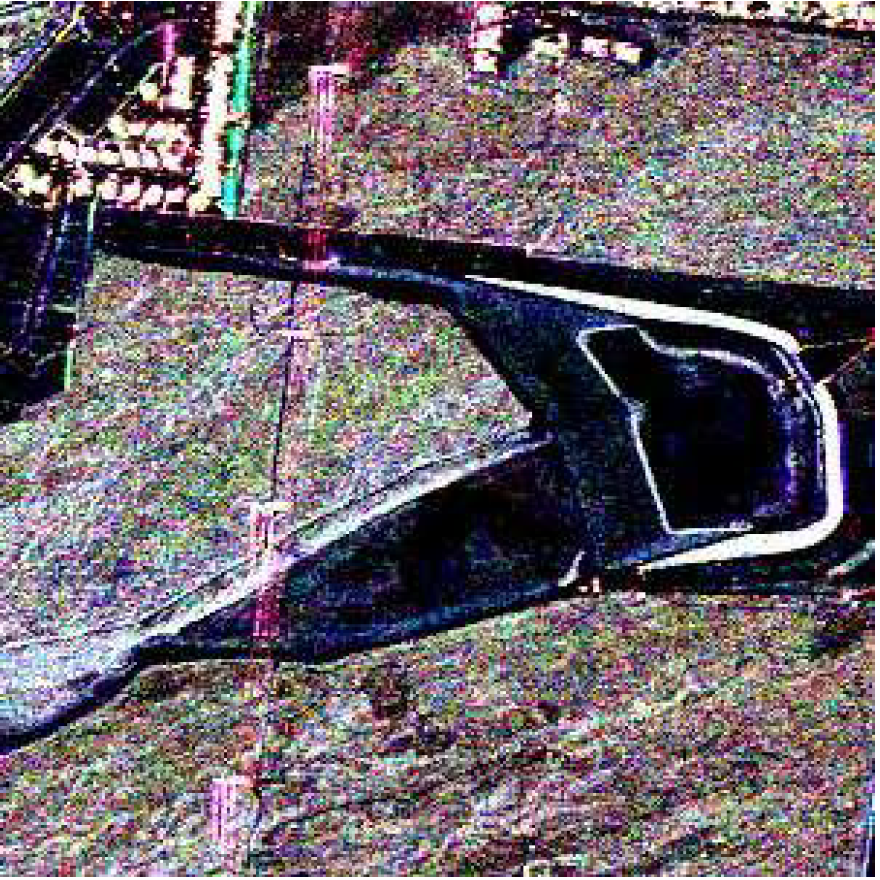}}
\subfigure[Scene 2 (after)\label{Application2:4}]{
\includegraphics[width=.43\linewidth]{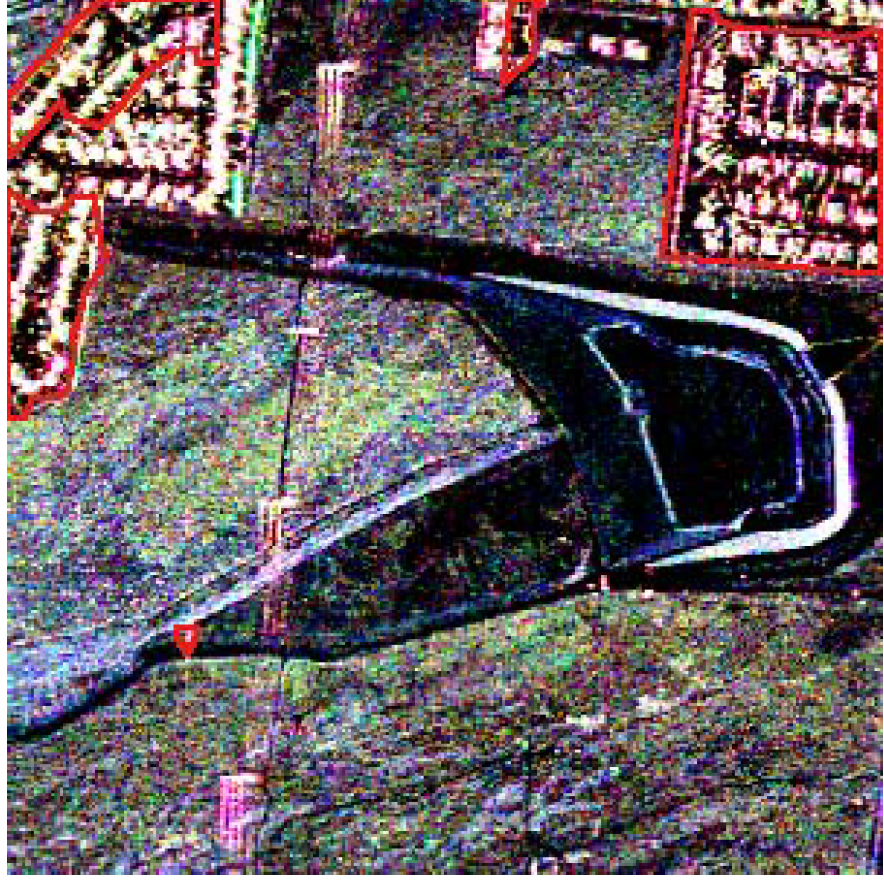}}
\subfigure[Scene 1 (reference map)\label{Application2:5}]{\fbox{
\includegraphics[width=.43\linewidth]{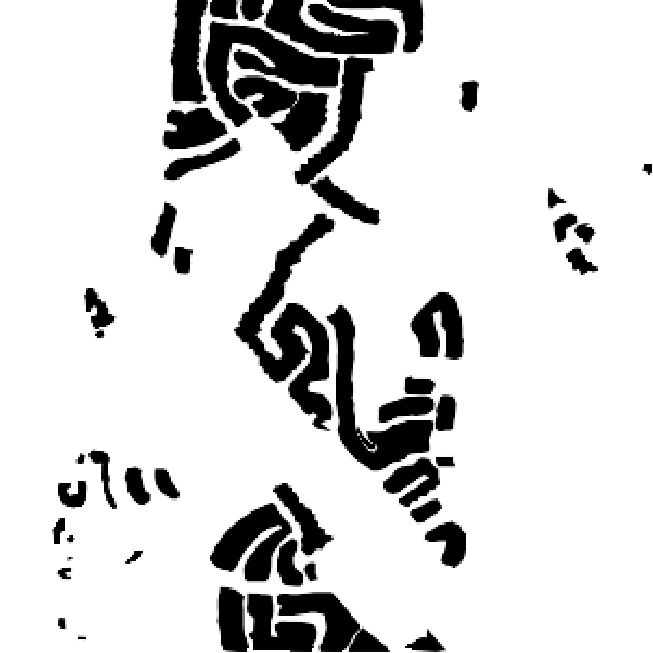}}
}
\subfigure[Scene 2 (reference map)\label{Application2:6}]{\fbox{
\includegraphics[width=.43\linewidth]{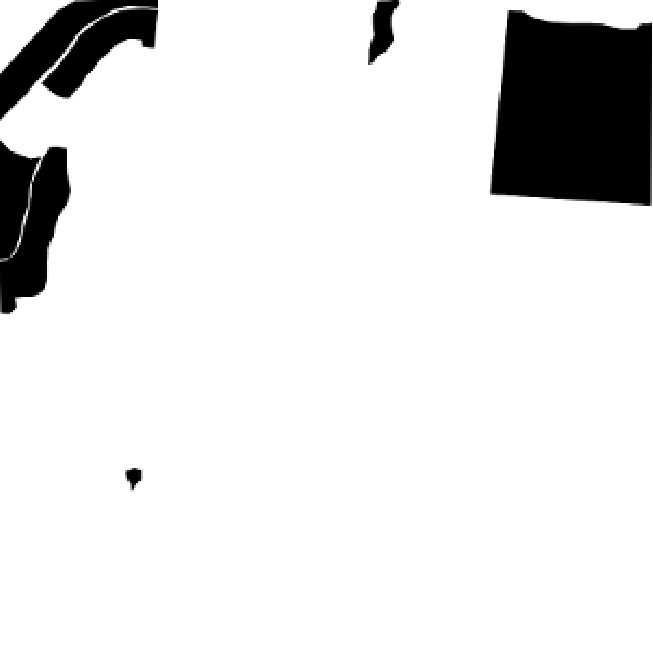}}
}
\caption{
UAVSAR images (in Pauli decomposition) 
on April 23, 2009 and May 11, 2015.
}
\label{Application2}
\end{figure}

Fig.~\ref{fig:StudyArea} presents the study areas for this experiment: 
surroundings of the city of Los Angeles, CA, USA.
These pictures refer to a dense urban area whose changes are caused by the urbanization process.
Ratha~\emph{et~al.}~\cite{Ratha2017} employed these data in the proposal of change detectors for single look polarimetric data using a geodesic distance.
Here we apply the four multilook PolSAR data detectors discussed in Section~\ref{comparison:methodology}.
Fig.~\ref{Application2} shows the Pauli decomposition of 
two UAVSAR images obtained by JPL's UAVSAR (\emph{Uninhabited Aerial Vehicle Synthetic Aperture Radar}) sensor at two different instants (23 April 2009, and 3 May 2015).

\begin{figure*}[hbt]
\centering
\subfigure[$S_{\text{R}}^{0.1}$-$3 \times 3$\label{Pvaluemaps1:1}]{
\includegraphics[width=.23\linewidth]{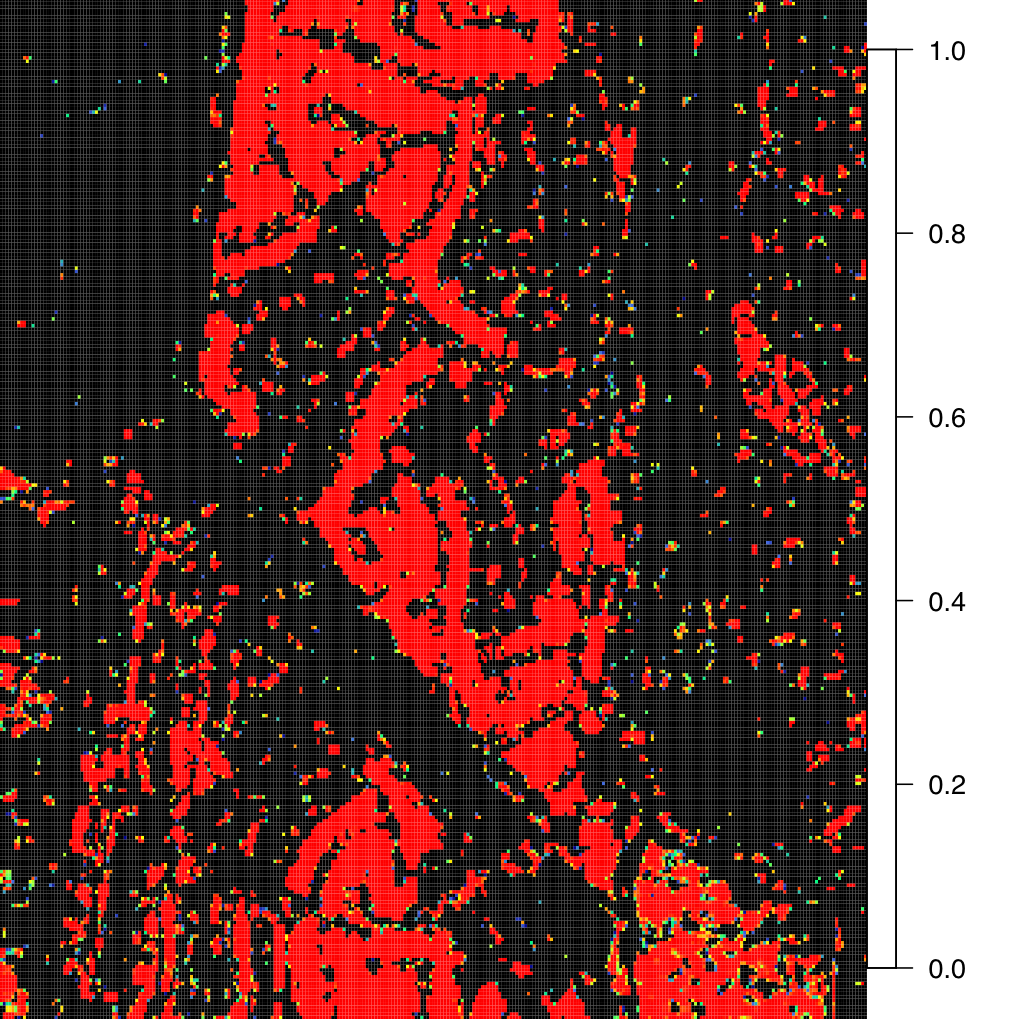}
}
\subfigure[$S_{\text{R}}^{0.1}$-$3 \times 3$\label{Pvaluemaps1:1:0}]{
\includegraphics[width=.23\linewidth]{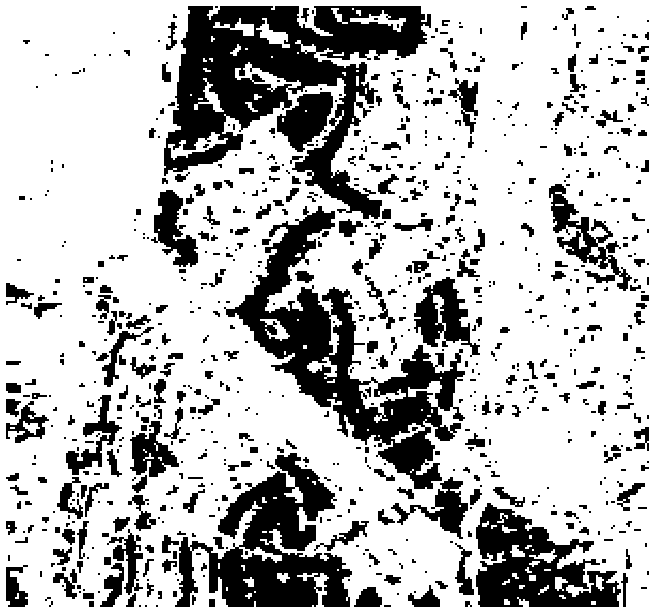}
}
\subfigure[$S_{\text{S}}$-$3 \times 3$\label{Pvaluemaps1:2}]{
\includegraphics[width=.23\linewidth]{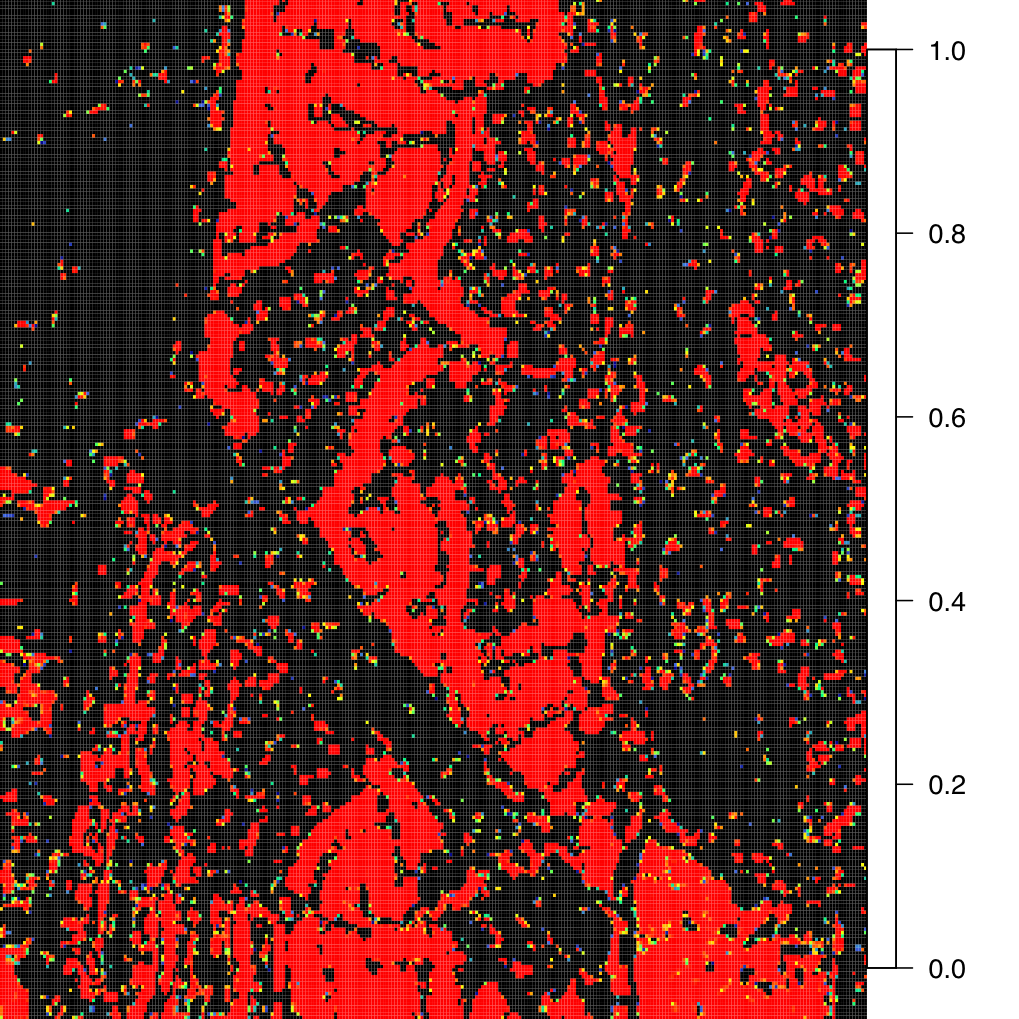}
}
\subfigure[$S_{\text{S}}$-$3 \times 3$\label{Pvaluemaps1:2:0}]{
\includegraphics[width=.23\linewidth]{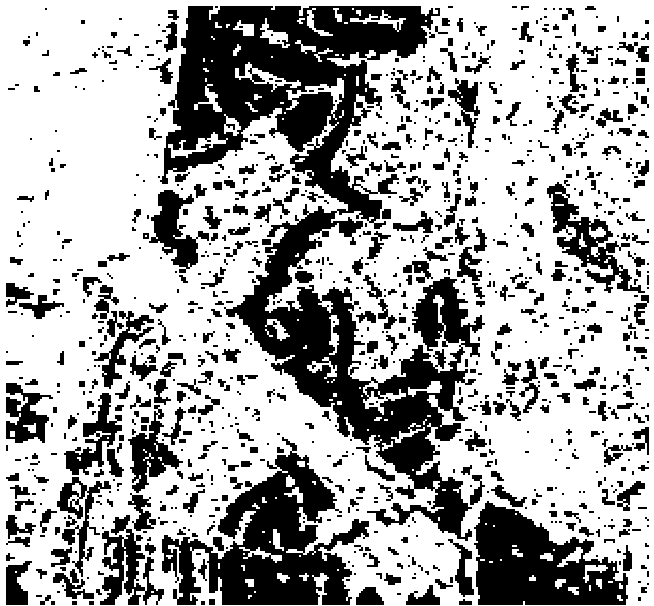}
}    
\\
\subfigure[$S_{\text{LR}}$-$3 \times 3$\label{Pvaluemaps1:3}]{
\includegraphics[width=.23\linewidth]{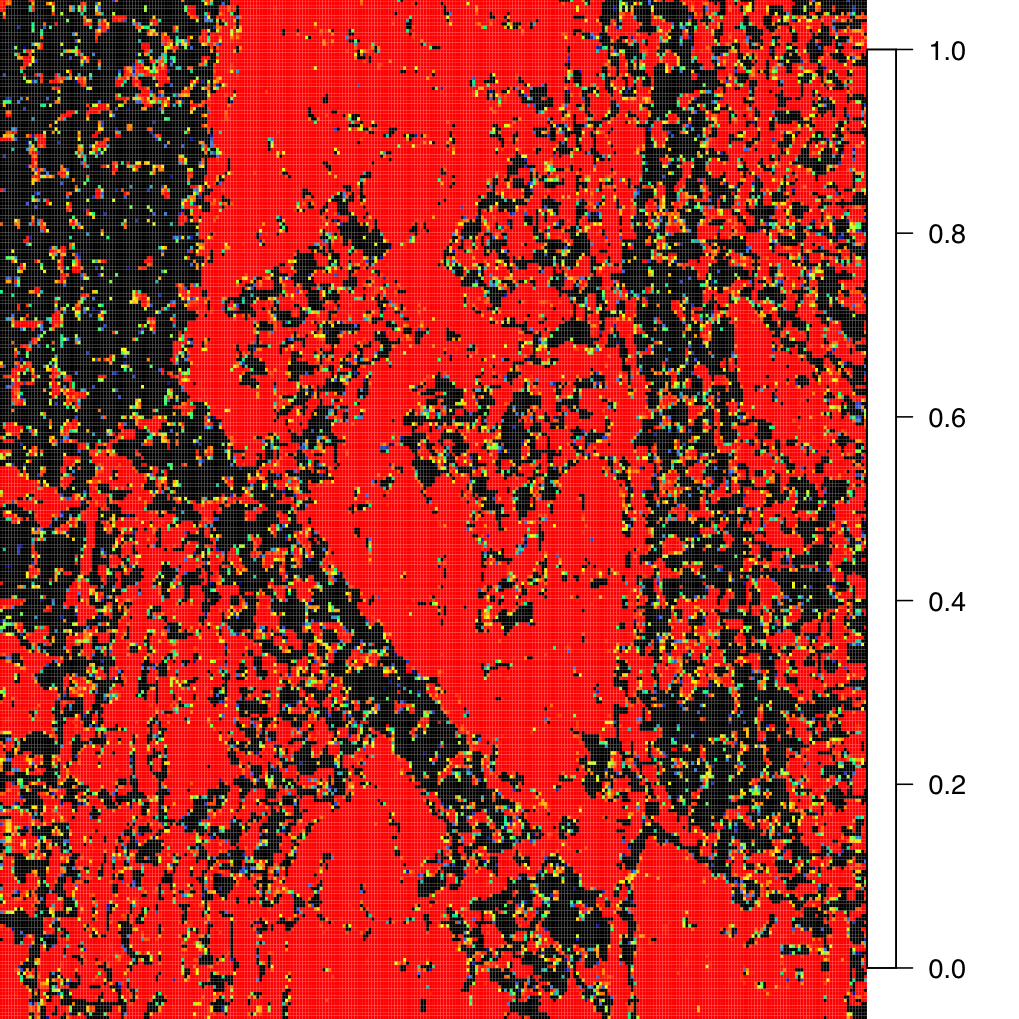}
}
\subfigure[$S_{\text{LR}}$-$3 \times 3$\label{Pvaluemaps1:3:0}]{
\includegraphics[width=.23\linewidth]{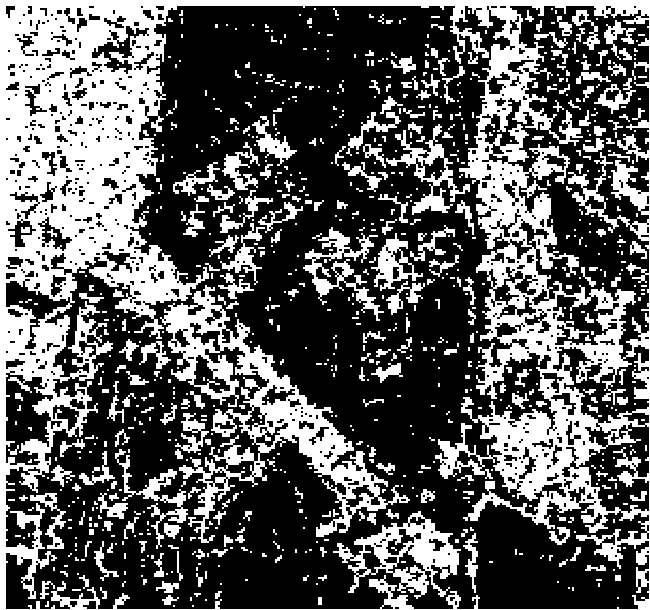}
}
\subfigure[$S_{\text{KL}}$-$3 \times 3$\label{Pvaluemaps1:4}]{
\includegraphics[width=.23\linewidth]{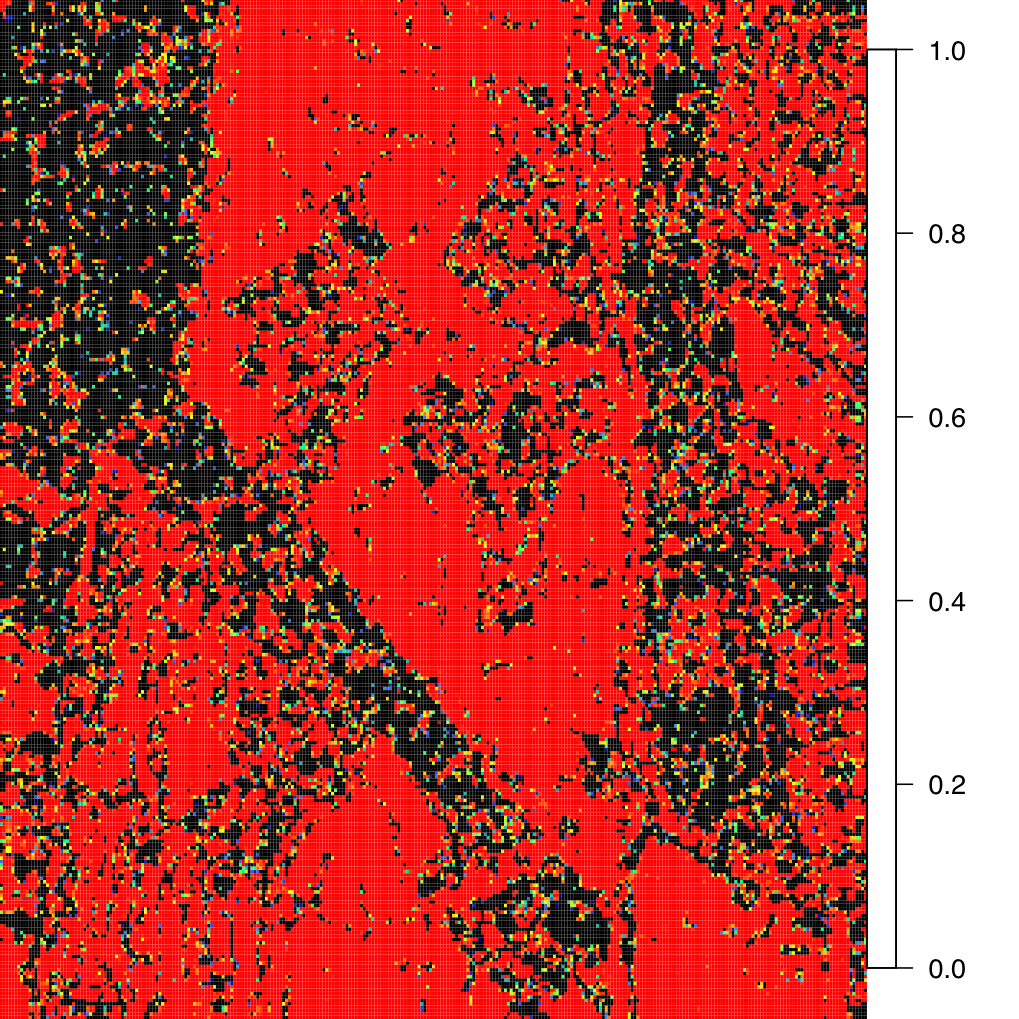}
}
\subfigure[$S_{\text{KL}}$-$3 \times 3$\label{Pvaluemaps1:4:0}]{
\includegraphics[width=.23\linewidth]{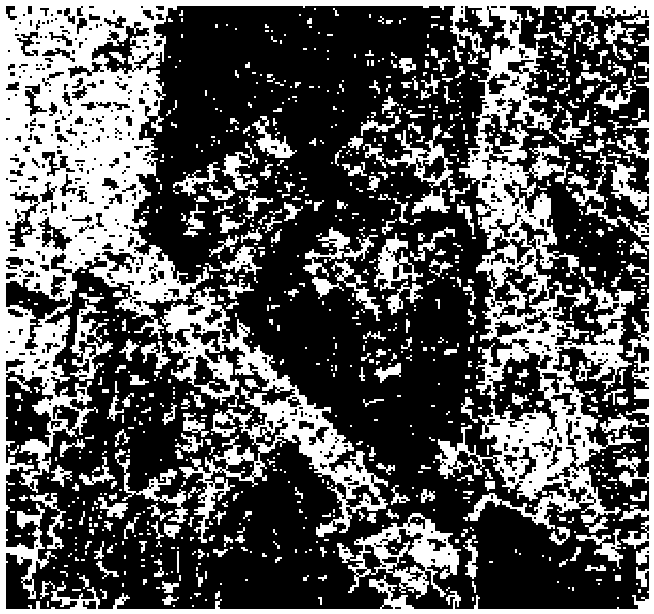}
}  
\caption{
$p$-value maps as evidence of changes between two dates for the first scene. 
}
\label{Pvaluemaps1}
\end{figure*}

\begin{figure*}[hbt]
\centering
\subfigure[$S_{\text{R}}^{0.1}$-$3 \times 3$\label{Pvaluemaps2:1}]{
\includegraphics[width=.23\linewidth]{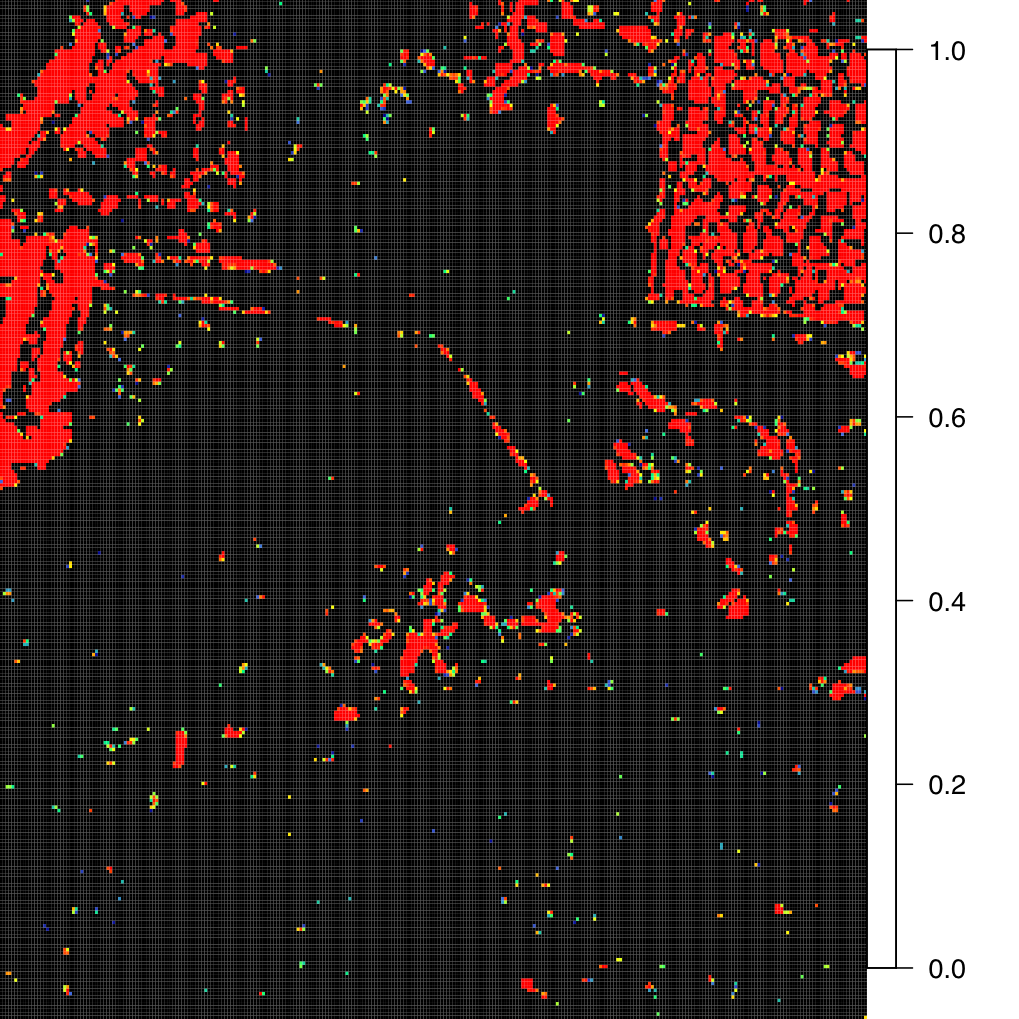}
}
\subfigure[$S_{\text{R}}^{0.1}$-$3 \times 3$\label{Pvaluemaps2:1:0}]{
\includegraphics[width=.23\linewidth]{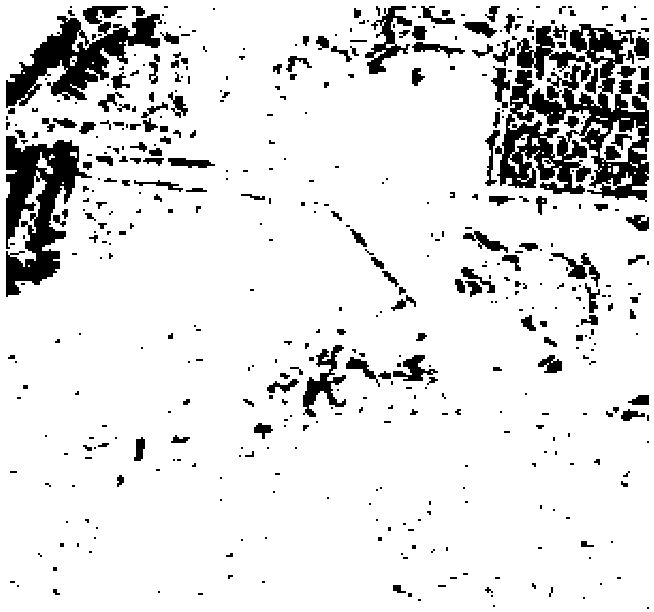}
}
\subfigure[$S_{\text{S}}$-$3 \times 3$\label{Pvaluemaps2:2}]{
\includegraphics[width=.23\linewidth]{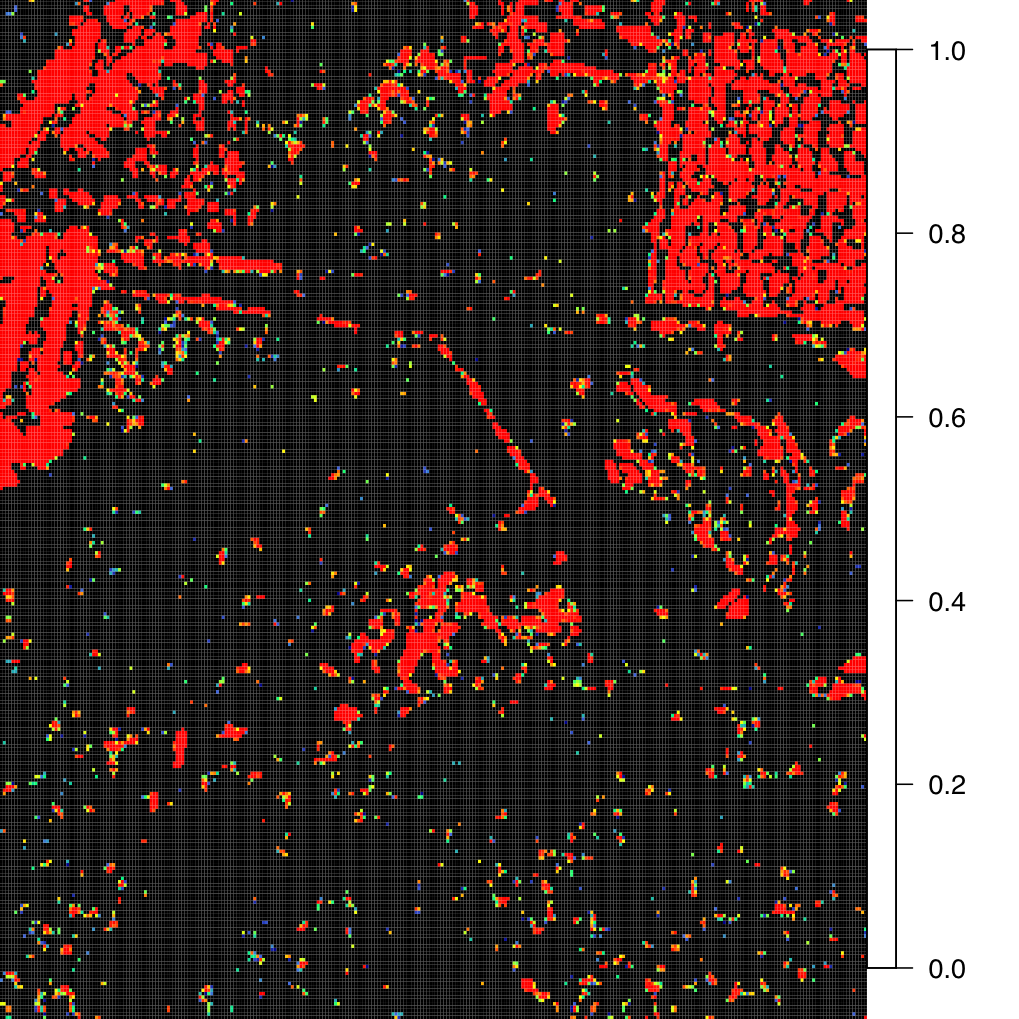}
}
\subfigure[$S_{\text{S}}$-$3 \times 3$\label{Pvaluemaps2:2:0}]{
\includegraphics[width=.23\linewidth]{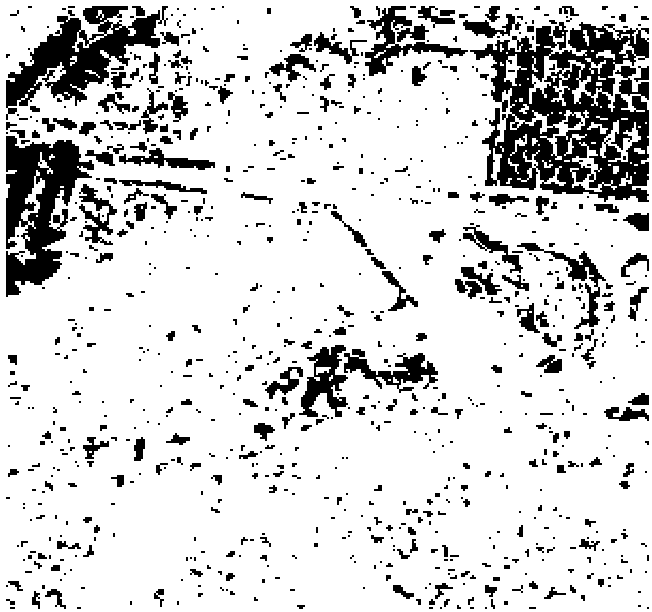}
}
\\
\subfigure[$S_{\text{LR}}$-$3 \times 3$\label{Pvaluemaps2:3}]{
\includegraphics[width=.23\linewidth]{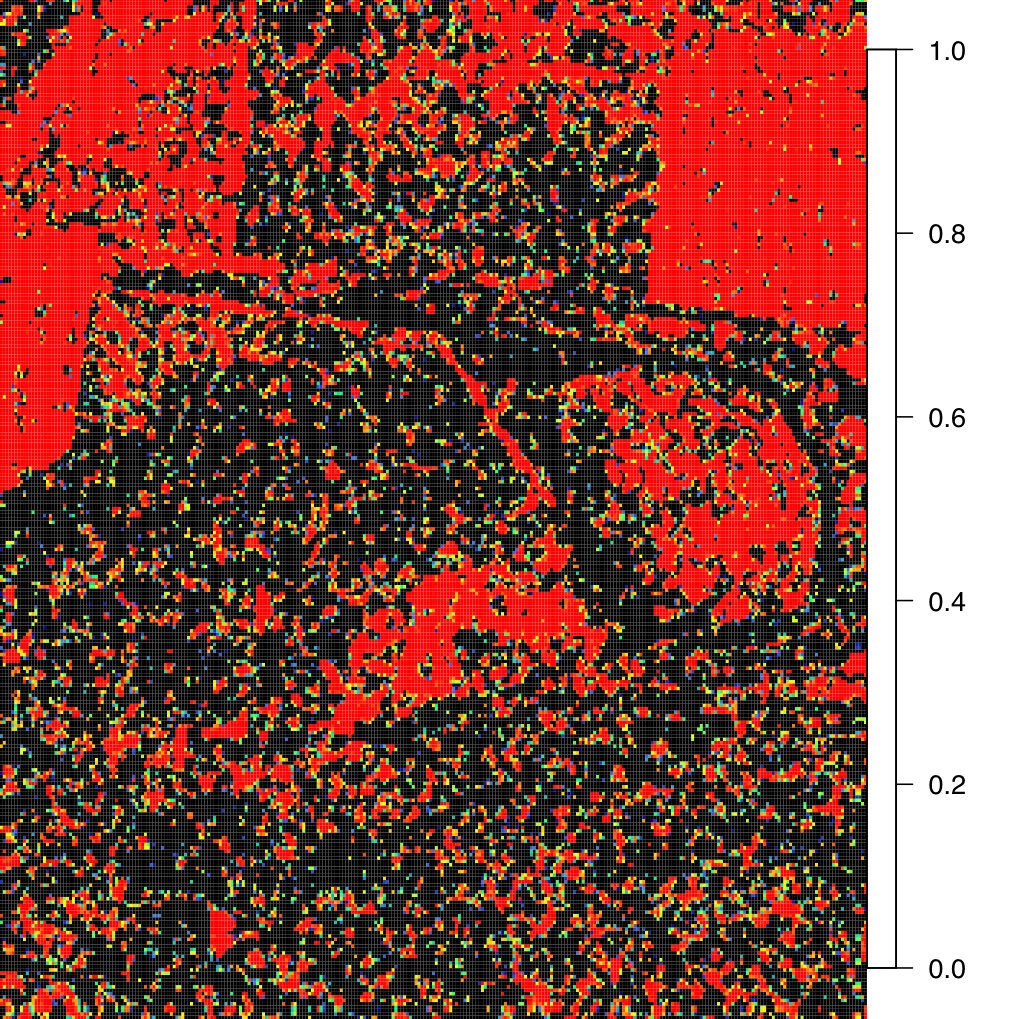}
}
\subfigure[$S_{\text{LR}}$-$3 \times 3$\label{Pvaluemaps2:3:0}]{
\includegraphics[width=.23\linewidth]{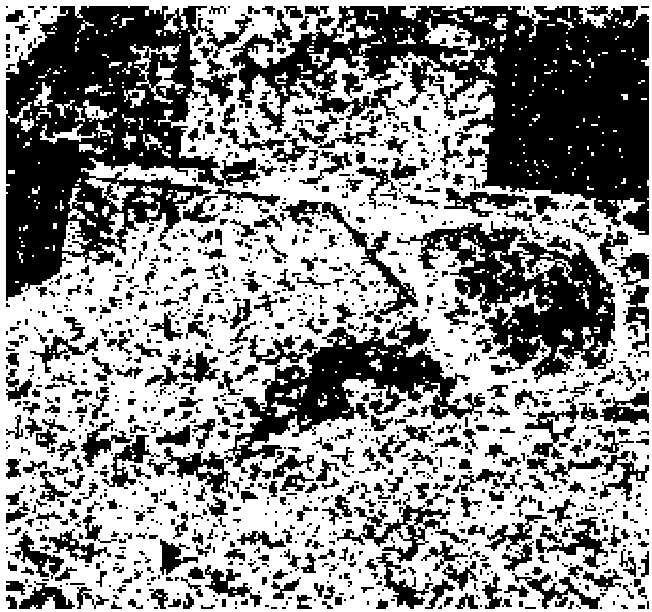}
}
\subfigure[$S_{\text{KL}}$-$3 \times 3$\label{Pvaluemaps2:4}]{
\includegraphics[width=.23\linewidth]{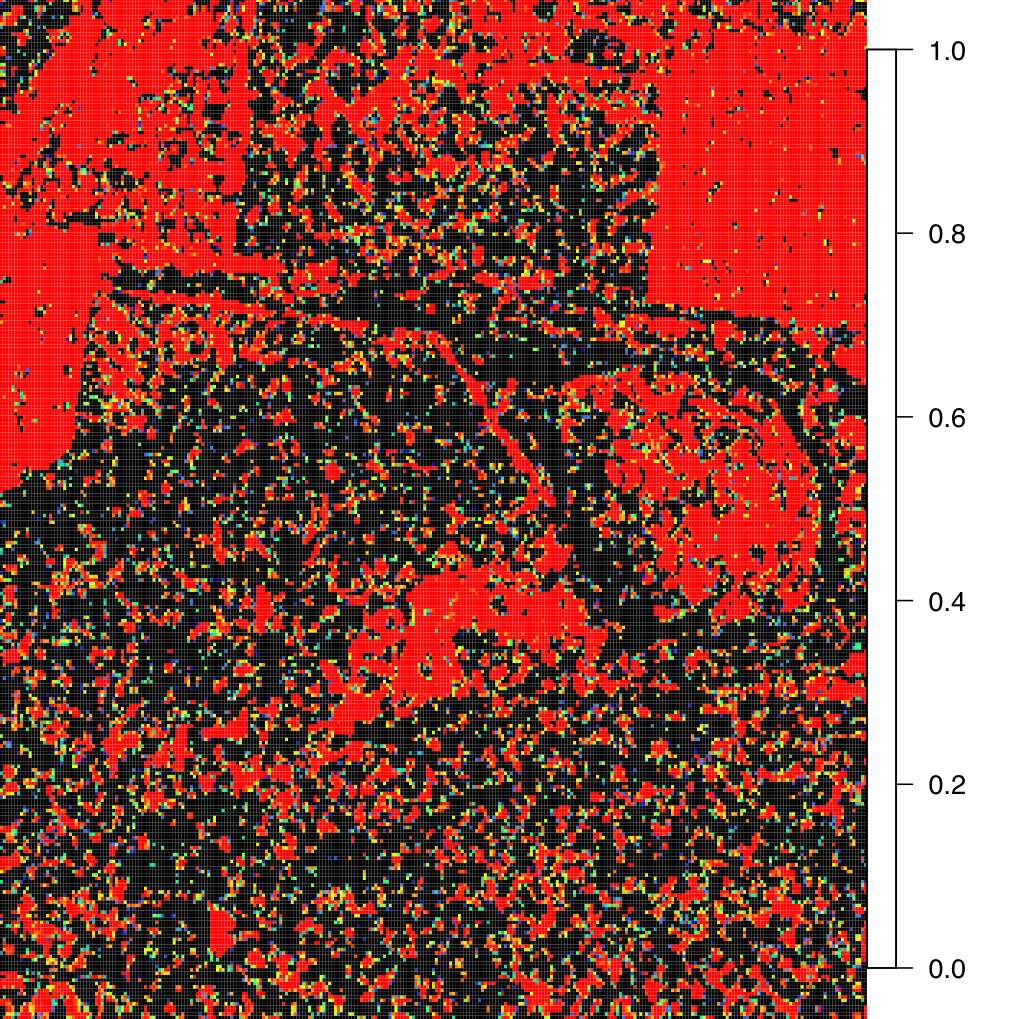}
}
\subfigure[$S_{\text{KL}}$-$3 \times 3$\label{Pvaluemaps2:4:0}]{
\includegraphics[width=.23\linewidth]{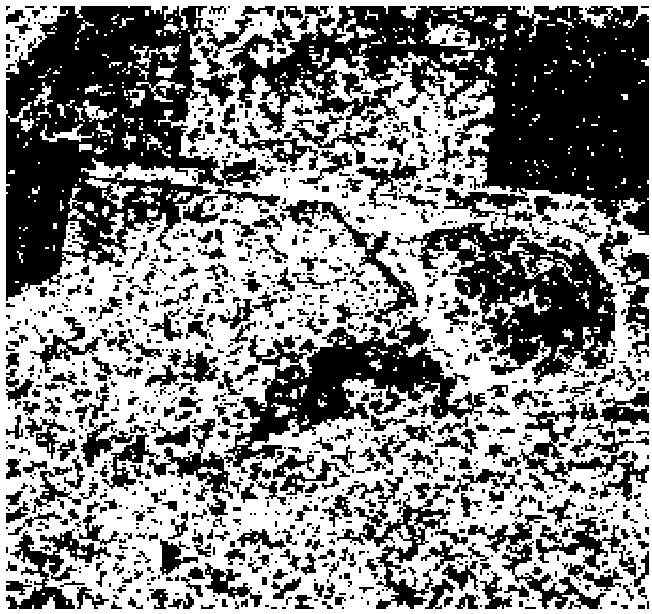}
}
\caption{
$p$-value maps as evidence of changes between two dates for the second scene. 
}
\label{Pvaluemaps2}
\end{figure*}

Using windows of size $3\times3$ on both dates, we computed the $S_{\text{LR}}$, $S_{\text{KL}}$, $S_{\text{S}}$, and $S_{\text{R}}^{0.1}$ test statistics and, from them, $p$-value maps; cf.\ Figs.~\ref{Pvaluemaps1} and~\ref{Pvaluemaps2}.
Probability values higher than \SI{0.01}{\percent} are drawn in black, as they provide no evidence of change.
Values below \SI{0.01}{\percent} range vary from red to dark blue (from strong to weak evidence of change).

It is noticeable
that $S_{\text{R}}^{0.1}$ and 
$S_{\text{S}}$ are similar,
cf.\ Figs.~\ref{Pvaluemaps1:1},~\ref{Pvaluemaps2:1} and~\ref{Pvaluemaps1:2},~\ref{Pvaluemaps2:2},
while $S_{\text{LR}}$ and $S_{\text{KL}}$ look alike, see Figs.~\ref{Pvaluemaps1:3}, \ref{Pvaluemaps2:3}, \ref{Pvaluemaps1:4}, and~\ref{Pvaluemaps2:4}, but somewhat different from the previous pair.

\begin{figure}[htb]
\centering
\includegraphics[width=.7\linewidth]{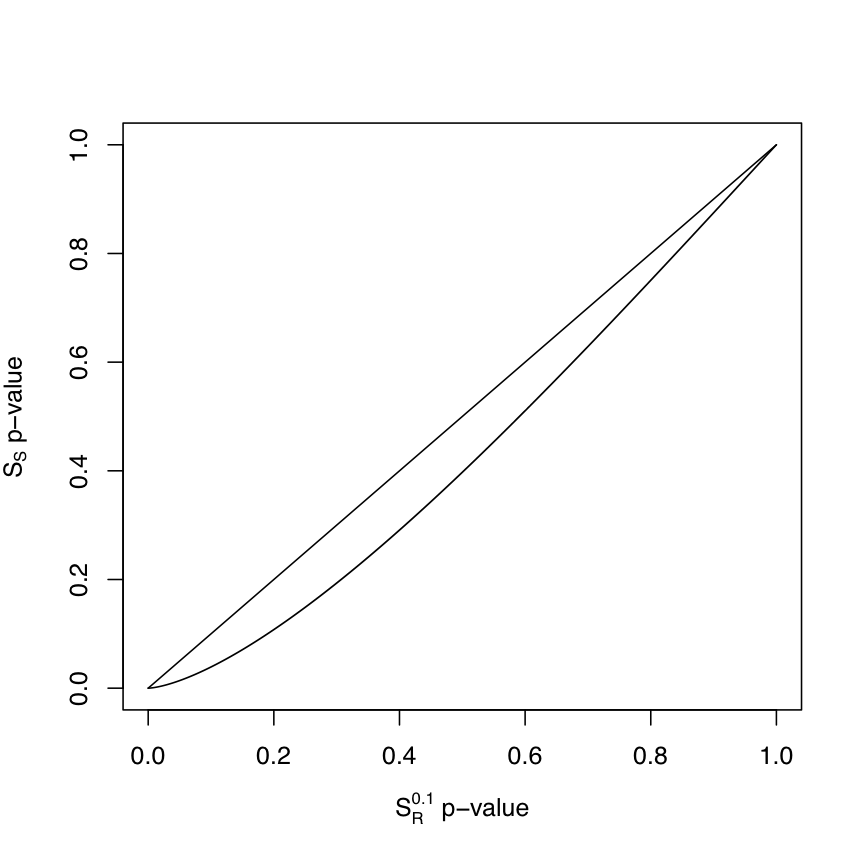}
\caption{Relationship between $p$-values from $S_{\text{R}}^{0.1}$ and $S_{\text{S}}$.}
\label{Rel_shannon_renyi}
\end{figure}

Fig.~\ref{Rel_shannon_renyi} shows the relationship between $S_{\text{R}}^{0.1}$ and $S_{\text{S}}$ for the second scene, along with the identity function for reference.
The $p$-values associated to the 
Shannon statistic are smaller than that those related to the R\'enyi statistic, so the former tends to reject more than $S_{\text{R}}^{0.1}$, as discussed in the simulation experiments.

Finally,
Figs.~\ref{Pvaluemaps1:1:0}, \ref{Pvaluemaps1:2:0}, \ref{Pvaluemaps1:3:0}, \ref{Pvaluemaps1:4:0}
and~\ref{Pvaluemaps2:1:0}, \ref{Pvaluemaps2:2:0}, \ref{Pvaluemaps2:3:0}, \ref{Pvaluemaps2:4:0}
show 
binary images resulting from thresholding 
the $S_{\text{R}}^{0.1}$, $S_{\text{S}}$, $S_{\text{LR}}$, and 
$S_{\text{KL}}$ statistics for the first and second scenes: $p$-values larger than $10^{-4}$ are shown in white, otherwise in black.
The results, again, favor entropy-based detectors.

To confirm the qualitative discussion, we quantify the performance of detectors with respect to reference maps in Figs~\ref{Application2:5}-\ref{Application2:6} in terms of five measures:
\begin{itemize}
\item False positive (FP):  Number of pixels indicated as change by Reference map (RM), but classified as no change; 
\item False negative (FN): Number of pixels indicated as no change by RM, but classified as change;
\item False alarm rate (FA): $(\text{FP}+\text{FN})/N$, where $N$ is the number of unchanged pixels according to the detector;
\item Detection rate (DR): $\text{TP}/\text{CG}$, where TP is the number of pixels indicated as change by both RM and the detector, and CG is the number of changed pixels according to the detector; and 
\item Kappa coefficient: $\kappa=(A-B)/(1-B)$, where 
$A=1-p_\text{FP}-p_\text{FN}$ and
$
B=(p_\text{TP}+p_\text{FP})(p_\text{TP}+p_\text{FN})+(p_\text{TN}+p_\text{FP})(p_\text{TN}+p_\text{FN}),
$
where $p_\mathcal{C}$ is the proportion of pixels under the condition $\mathcal{C}$
relative to the total number of pixels
and TN is the number of pixels indicated as no change by both RM and the detector.
\end{itemize}
The reference maps were prepared by specialists with Bing and Google Earth imagery; cf.\ Ref.~\cite{Ratha2017}.

Table~\ref{quantitative:analysis} shows the results.
$S_{\text{R}}$ obtained the best performance, followed by $S_{\text{S}}$, for both data sets with respect to $\kappa$ and DR.
These detectors presented lower FN and FA than $S_{\text{LR}}$ and $S_{\text{KL}}$.
$S_{\text{KL}}$ and $S_{\text{LR}}$ performed better than entropy-based detectors with respect to FP.
The values of FP were smaller than \SI{5}{\percent} in all cases, so this is not an issue for any detector. 

\begin{table}
\centering
\caption{
Detectors performance 
}
\label{quantitative:analysis}
\begin{tabular}{c|ccccc}
\toprule
Detectors  &  FP (\%)     &   FN (\%) &   FA (\%)  &   DR(\%)  &  $\kappa$ (\%)
\\ \midrule
& \multicolumn{5}{c}{ Scene 1 }
\\ \midrule
%
$S_{\text{LR}}$  & 0.060 &  13.433 & 13.493 & 20.408 & 24.597 \\
$S_{\text{KL}}$  & 0.052 &  14.055 & 14.107 & 19.765 & 23.023 \\
$S_{\text{S}}$   & 0.343 &   5.476 &  5.819 & 35.387 & 52.482 \\
$S_{\text{R}}$   & 0.431 &   3.709 &  4.140 & 42.988 & 62.272 
\\ \midrule
& \multicolumn{5}{c}{ Scene 2 }
\\ \midrule
%
%
$S_{\text{LR}}$  & 0.104 &    9.551 &  9.655 & 20.964 & 25.552 \\
$S_{\text{KL}}$  & 0.094 &   10.404 & 10.497 & 19.682 & 23.087 \\
$S_{\text{S}}$   & 0.739 &    2.568 &  3.307 & 41.598 & 52.598 \\
$S_{\text{R}}$   & 0.920 &    1.254 &  2.174 & 55.590 & 62.519
\\ \bottomrule
\end{tabular}
\end{table}

\section{Conclusions}\label{comparison:conclusion}

We quantified and compared the performance of four change detection methods for fully polarimetric SAR data. 
These methods are based on the likelihood-ratio statistic, on the Kullback-Leibler distance, and on the R\'enyi and Shannon entropies. 
We used empirical test powers and sizes as comparison criteria.

Firstly, the performance of the methods was quantified through a Monte Carlo study using scenarios modeled by the scaled complex Wishart law.
The empirical test sizes showed evidence that the detectors based on the likelihood ratio and Shannon entropy statistics presented the best performance.
In particular, the one based on the entropy is the best for small samples and statistically similar to the $S_{\text{LR}}$.
Additionally, the tests based on the Kullback-Leibler and on the likelihood ratio statistics tend to overestimate the nominal level, while those which employ entropies underestimate it.

Regarding the empirical test power, the test based on the Shannon entropy presented, in a consistent fashion, the best results.
Computational costs are quite different. 
The test statistic based on the likelihood ratio $S_{\text{LR}}$ requires evaluating~\eqref{SRex}, 
while $S_{\text{KL}}$ depends only on the Kullback-Leibler distance~\eqref{expreKL}.
The latter is less demanding than the former by an order of magnitude.
Thus, on those situations in which $S_{\text{KL}}$ and $S_{\text{LR}}$ are competitive (for moderate and large sample sizes), the Kullback-Leibler test is more attractive because it has the lowest computational cost.

Secondly, and since estimated test sizes were quite competitive, two experiments with actual data were performed.
For the single date experiment, in all the situations considered, the test based on R\'enyi entropy with order $\beta=0.1$ presented the best results.
The multitemporal data experiments suggests that change detectors equipped by entropies provide better performance than those based on the Kullback-Leibler distance and those based on the likelihood ratio statistic.
Finally, the diversity of tests statistics stemming from Information Theory opens the venue for investigation of composite decision rules, as in Ref.~\cite{OpticalBasedSAREdgeDetectionKNOWSys}.

Future works  will aim to adapt developments made in this paper to more general distributions as, for instance, the $\mathcal G_{\text{Pol}}$ and its particular cases ($\mathcal K_{\text{Pol}}$, $\mathcal G^0_{\text{Pol}}$, and $\mathcal G^H_{\text{Pol}}$), the Kummer-$\mathcal U$, and $\mathcal M$ laws; cf.~\cite[section~4.1]{SurveyStatisticalPolSAR}, and~\cite{PolarimetricSegmentationBSplinesMSSP}.

\appendix

\section{
Derivation outline of expressions for Shannon and R\'enyi statistics
}
\label{AP}

Applying~\eqref{eq:denswishart} in~\eqref{entropy:1} and~\eqref{entropy:2}, 
we obtain the following entropies~\cite{FreryCintraNascimento2012}:
\begin{align}
H_\text{S}&(\boldsymbol{\theta})=\frac{p(p-1)}{2}\log \pi- p^2 \log L +p \log |\boldsymbol{\Sigma}| + p L\nonumber \\
&\mbox{}+(p-L)\psi_p^{(0)}(L) +\displaystyle \sum_{k=0}^{p-1}\log \Gamma(L-k), \text{ and}\\
H_\text{R}^{\beta}&(\boldsymbol{\theta})= \frac{p(p-1)}{2}\log\pi- p^2 \log L +p \log |\boldsymbol{\Sigma}|\nonumber \\
&-\frac{pq\log\beta}{1-\beta}+\frac{\sum_{i=0}^{p-1}\bigr[\log\Gamma(q-i)-\beta\log\Gamma(L-i)\bigl]}{1-\beta},
\label{renyiw}
\end{align}
where $q=L+(1-\beta)(p-L)$.

Under the scaled complex Wishart law, Frery~\emph{et~al.}~\cite{FreryCintraNascimento2012} derived the following variances:
\begin{itemize}
\item Shannon:
\begin{align}
\sigma_{\text{S}}^2=&\frac{\bigl[(p-L)\psi_p^{(1)}(L)+p-\frac{p^2}{L}\bigr]^2}{\psi_p^{(1)}(L) - \frac{p}{L}}\nonumber \\
&+\frac{p^2}{L} \operatorname{vec}\bigl(\boldsymbol{\Sigma}^{-1}\bigr)^* \bigl( \boldsymbol{\Sigma} \otimes \boldsymbol{\Sigma}\bigr)  \operatorname{vec}\bigl(\boldsymbol{\Sigma}^{-1}\bigr).
\label{varianceentropyshannon}
\end{align}
\item R\'enyi entropy:
\begin{align}
\sigma_{\text{R},\beta}^2=&
\frac{\Big\{\frac{\beta}{1-\beta} \bigl[\psi_p^{(0)}(q)-\psi_p^{(0)}(L)\bigr]-\frac{p\beta\ln(\beta)}{1-\beta}-\frac{p^2}{L}\Big\}^2}{\psi_p^{(1)}(L) - \frac{p}{L}}\nonumber \\
&+\frac{p^2}{L} \operatorname{vec}\bigl(\boldsymbol{\Sigma}^{-1}\bigr)^* \bigl(\boldsymbol{\Sigma} \otimes \boldsymbol{\Sigma}\bigr)\operatorname{vec}\bigl(\boldsymbol{\Sigma}^{-1}\bigr).
\label{varianceentropyrenyi}
\end{align}
\end{itemize}		

\bibliographystyle{IEEEtran}
\bibliography{bibtexart}

\begin{IEEEbiography}[{\includegraphics[width=1in]{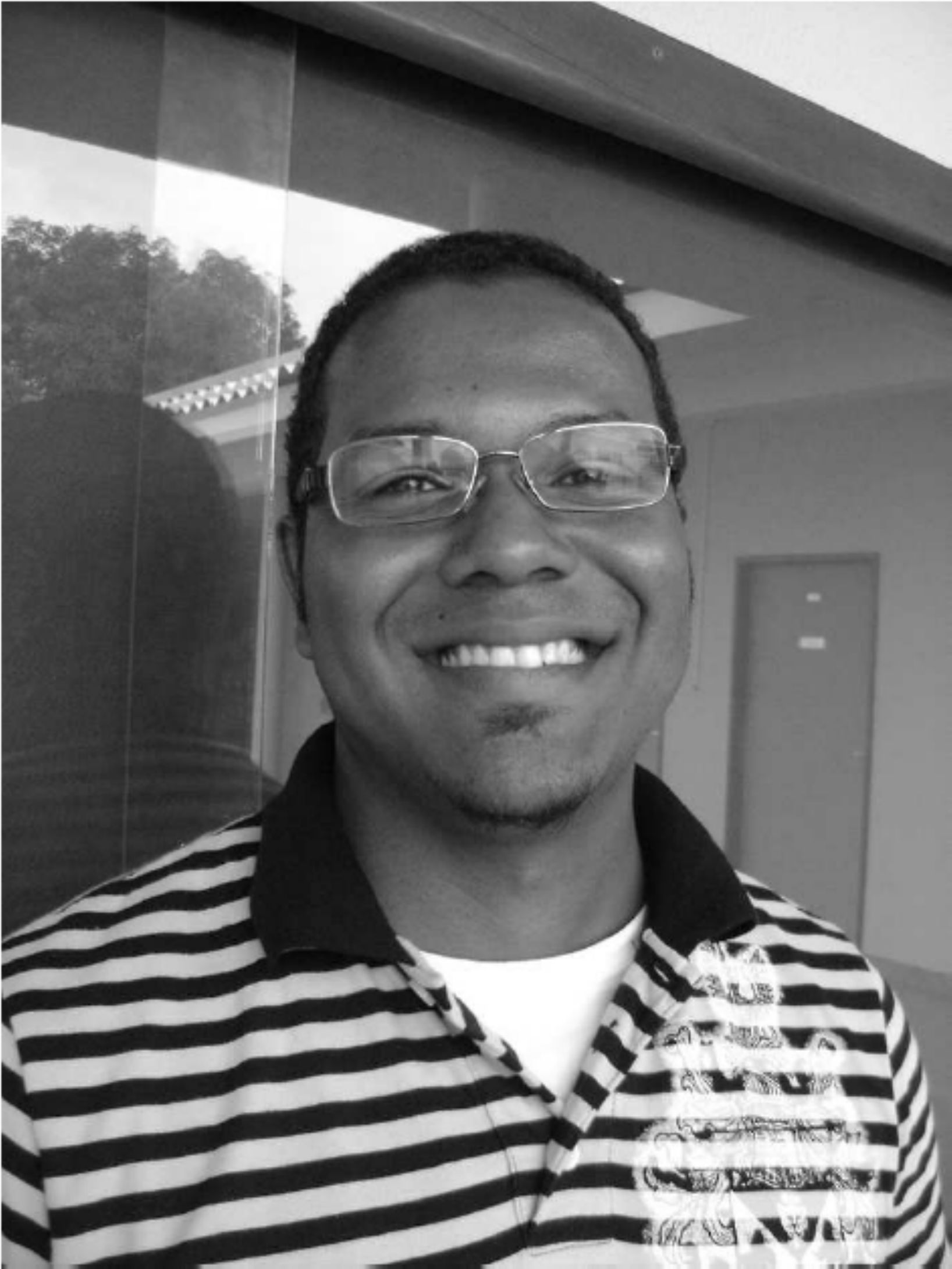}}]{Abra\~ao D.\ C.\ Nascimento}
holds B.Sc.\, M.Sc.\, and D.Sc. degrees in Statistics from Universidade Federal de Pernambuco (UFPE), Brazil, in 2005, 2007, and 2012, respectively.
In 2014, he joined the Department of Statistics at UFPE as Adjoint Professor.
His research interests are statistical information theory, inference on random matrices (with emphasis for applications on PolSAR imagery), statistical theory of shape, spatio-temporal processes, survival analysis, and asymptotic theory.
\end{IEEEbiography}

\begin{IEEEbiography}[{\includegraphics[width=1in]{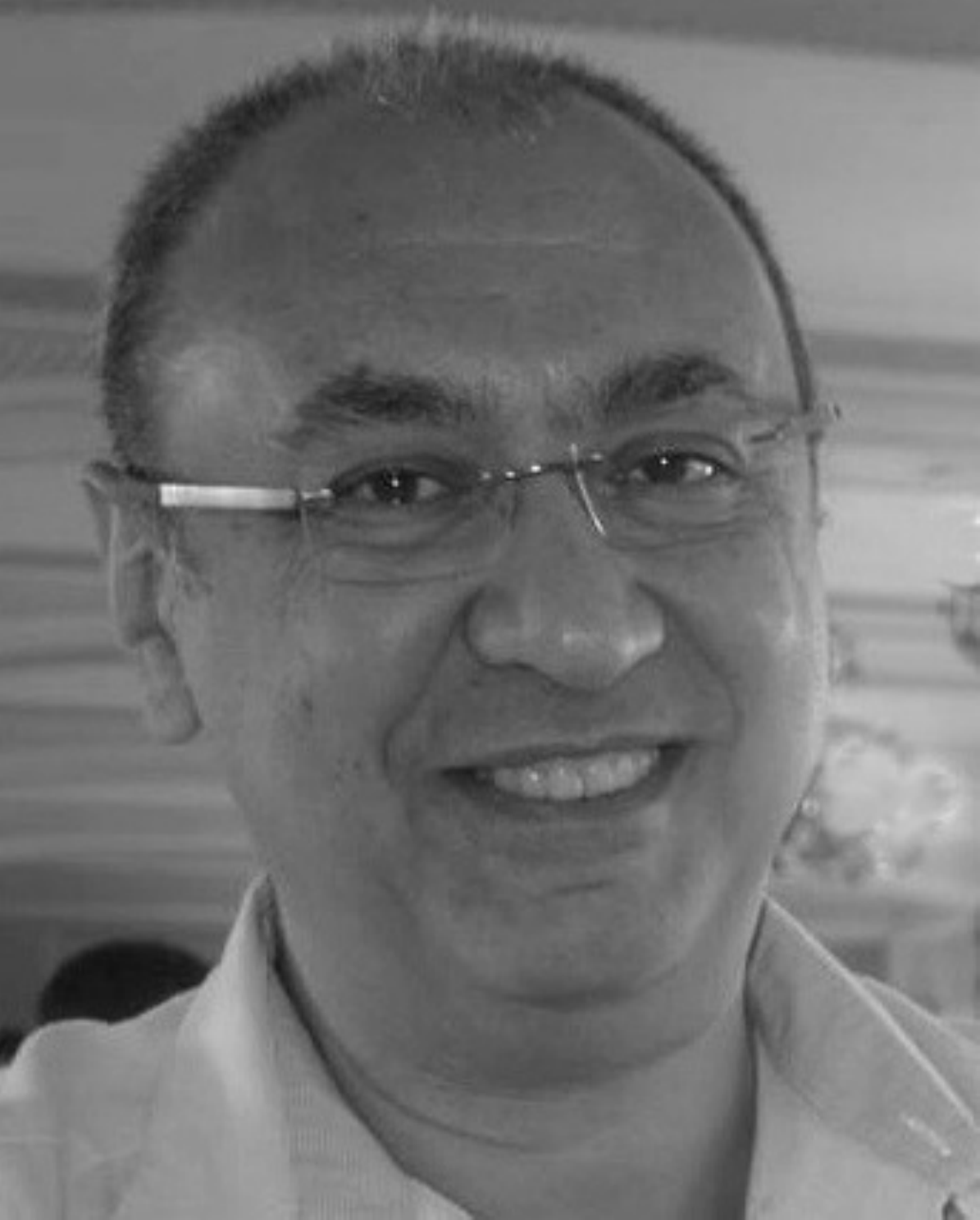}}]{Alejandro C.\ Frery} (S'92--SM'03)
received the B.Sc. degree in Electronic and Electrical Engineering from the Universidad de Mendoza, Mendoza, Argentina.
His M.Sc. degree was in Applied Mathematics (Statistics) from the Instituto de Matem\'atica Pura e Aplicada (IMPA, Rio de Janeiro) and his Ph.D. degree was in Applied Computing from the Instituto Nacional de Pesquisas Espaciais (INPE, S\~ao Jos\'e dos Campos, Brazil).
He is currently the leader of LaCCAN -- \textit{Laborat\'orio de Computa\c c\~ao Cient\'ifica e An\'alise Num\'erica}, Universidade Federal de Alagoas, Macei\'o, Brazil.
His research interests are statistical computing and stochastic modelling.
\end{IEEEbiography}

\begin{IEEEbiography}[{\includegraphics[width=1in]{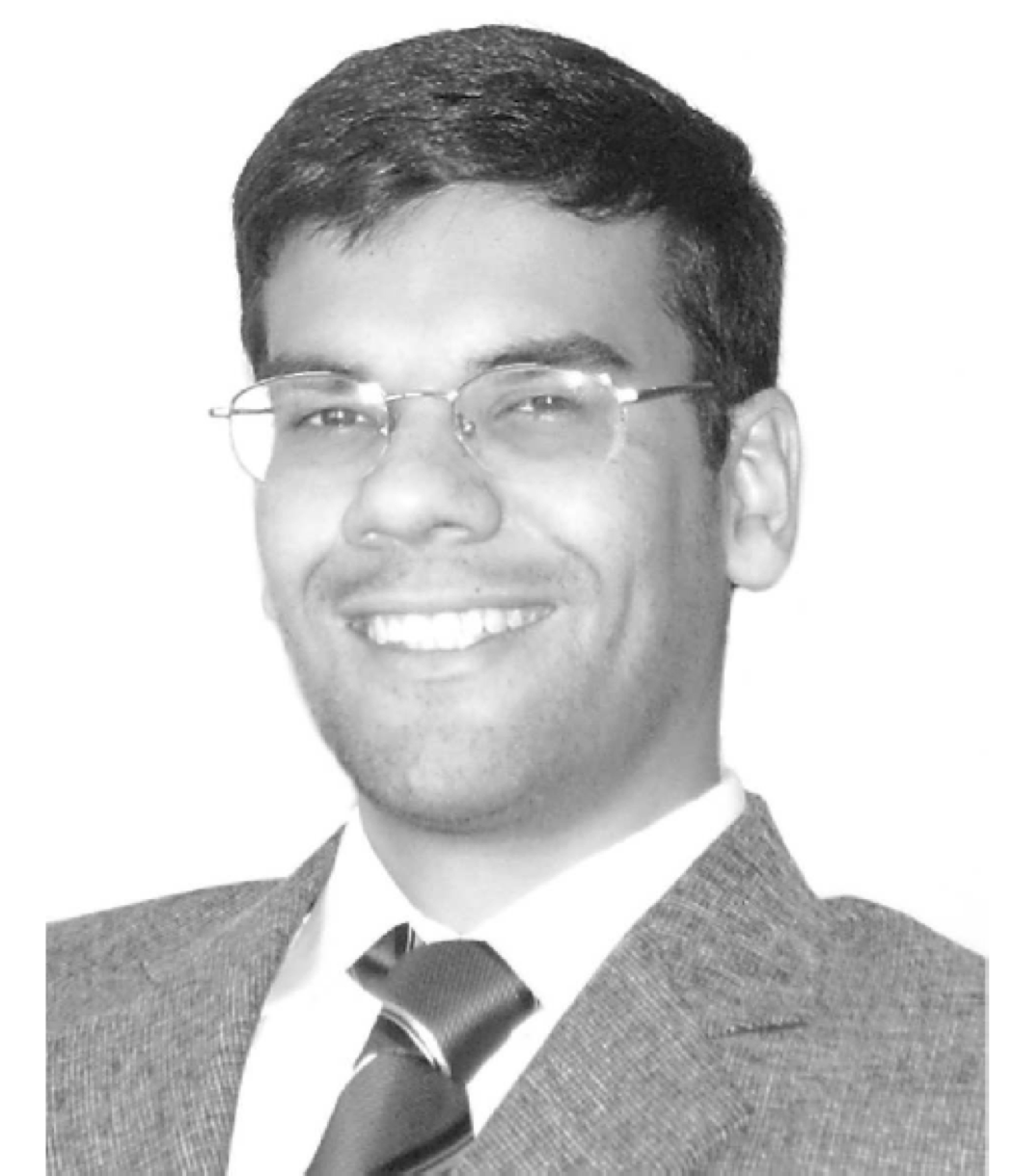}}]{Renato~J.~Cintra} (S'00, M'05, SM'10)
received the B.Sc., M.Sc., and D.Sc. degrees in electrical engineering from the Universidade Federal de Pernambuco (UFPE), Recife, Brazil, in 1999, 2001, and 2005, respectively. He joined the Department of Statistics, UFPE, in 2005. In 2014--2015, he was visiting professor at the D\'epartement Informatique, INSA, Lyon, France. During 2017--2018, he is a visiting professor at the University of Calgary, Canada. He is an associate editor for IEEE Geoscience and Remote Sensing Letters; Springer Circuits, Systems, and Signal Processing; IET Circuits, Devices \& Systems; and Journal of Communication and Information Systems. His long-term topics of research include: approximation theory for discrete transforms, theory and methods for digital signal processing, and
statistical methods.
\end{IEEEbiography}
\vfill

\end{document}